\documentclass[12pt,draft]{article}

\usepackage{mathrsfs,bm,color,sasaki,braket}
\usepackage{amsmath,amssymb,amsthm}
\usepackage[a4paper]{geometry}
\newgeometry{left=30mm,right=30mm,bottom=32mm,top=32mm}

\theoremstyle{plain}
\newtheorem{theorem}{Theorem}[section]
\newtheorem{df}[theorem]{\bf Definition}
\newtheorem{thm}[theorem]{\bf Theorem}
\newtheorem{cor}[theorem]{\bf Corollary}
\newtheorem{lem}[theorem]{\bf Lemma}

\newtheorem{prop}[theorem]{\bf Proposition}

\numberwithin{equation}{section}

\theoremstyle{remark}
\newtheorem{rem}[theorem]{\bf Remark}
\newcommand{\phih}{\hat\varphi}
\newcommand{\phiome}{\phi_\omega}
\newcommand{\Ffin}{\mathscr{F}_\mathrm{fin}}
\newcommand{\Hfin}{\mathscr{H}_\mathrm{fin}}
\newcommand{\Hfm}{H_{\mathrm{f},m}}
\newcommand{\Hf}{H_\mathrm{f}}
\newcommand{\sF}{\mathscr{F}}
\newcommand{\sH}{\mathscr{H}}

\newcommand{\Czinf}{C_0^\infty(\RR^3)}
\newcommand{\half}{\frac{1}{2}}
\renewcommand{\dom}{\mathrm{D}}

\newcommand{\stensor}{\tensor_\mathrm{s}}
\def\bbbone{{\mathchoice {\rm 1\mskip-4mu l} {\rm 1\mskip-4mu l}
{\rm 1\mskip-4.5mu l} {\rm 1\mskip-5mu l}}}
\def\one{\bbbone}

\newcommand{\cP}{\mathcal{P}}
\newcommand{\Nf}{{\rm N}}
\newcommand{\hi}{H_{\rm I}}
\newtheorem{exa}[theorem]{\bf Example}

\title{\sc Spectrum of the semi-relativistic Pauli-Fierz model II}

\author{
Takeru Hidaka\thanks{Faculty of Mathematics, Kyushu university, Fukuoka 819-0395, Japan}, 
Fumio Hiroshima\thanks{Faculty of Mathematics, Kyushu university, Fukuoka 819-0395, Japan} and 
Itaru Sasaki\thanks{Department of Mathematics, Shinshu university, Matsumoto 390-8621, Japan}
}
\date{\today}
\begin{document}
\maketitle
%%%%%%%%%%%%%%%% end

%%%%%%%%%%%%%%%% abstract
\begin{abstract}
We consider the semi-relativistic Pauli-Fierz Hamiltonian
\begin{align*}
H_m = |\bp-\bA(\bx)| + \Hfm + V(\bx),\quad m\geq0,
\end{align*}
and prove the existence of the ground state of $H_m$ for $m=0$. 
Here $\bA(\bx)$ denotes a quantized radiation field and 
$\Hfm$ the free field Hamiltonian 
with the dispersion relation $\sqrt{|\bk|^2+m^2}$ with $m\geq0$. 
This paper is the sequel of \cite{hh16}, where 
the existence of the ground state $\Phi_m$ of $H_m$ for $m>0$ is proven.  
In order to show the existence of the ground state  for $m=0$ 
we estimate a singular and non-local pull-through formula and 
show the equicontinuity of set $\{a(k)\Phi_m\}_{0<m<m_0}$ with some $m_0$, where $a(k)$ denotes 
the formal kernel of the annihilation operator. 
Taking a subsequence $m_j$, we can conclude that $\lim_{m_j\to0}\Phi_{m_j}=\Phi_0\not=0$ and $\Phi_0$ is the ground state of $H_0$.
\end{abstract}
%%%%%%%%%%%%%%%% end

%%%%%%%%%%%%%%%% introduction
\section{Introduction}
\subsection{Semi-relativistic Pauli-Fierz model}
In this paper, we are concerned with the existence of the ground state of the 
so-called semi-relativistic Pauli-Fierz  model (it is abbreviated as SRPF model), 
which describes an interaction 
between a semi-relativistic charged particle and a quantized radiation  field.
To show the existence of a  ground state in quantum field theory has been a fascinating problem, and  
the existence of the ground state of typical models including 
the non-relativistic Pauli-Fierz mode \cite{pf38}, the SRPF model with a massive particle, the Nelson mode \cite{ne64} and spin-boson model  has been proven.   
As far as we know, however, that  of the SRPF model with a massless particle has been left so far. 

The non-relativistic Pauli-Fierz Hamiltonian is given by 
$$H_{\rm PF}=\frac{1}{2M}(\bp-\bA(\bx))^2+\Hfm + V(\bx),$$
where $M$ denotes the mass of a charged particle, $\bp$ the $3$-dimensional momentum operator,   
$\bA(\bx)=\bA_{\hat\varphi}(\bx)$ a quantized radiation field with an ultraviolet cutoff function $\hat \varphi$, 
$\Hfm$ the free field Hamiltonian with dispersion relation 
$\omega_m(\bk)=\sqrt{|\bk|^2+m^2}$ with photon mass $m\geq0$ and photon momentum $\bk\in\RR^3$, 
and $V(\bx)$ an external potential. 
The spectrum of $H_{\rm PF}$ has been studied in e.g. \cite{bfs3,gll,ll03} 
as well as the Nelson model in e.g., \cite{bfs1,bfs2,sp98,ge00} and 
spin-boson model in e.g., \cite{sp89,ah97}. 
%It is comprehensively reviewed in e.g. \cite{hir19}. 
The existence and uniqueness of the ground state of $H_{\rm PF}$ is established for $m\geq0$ under some conditions on $V$ and  $\hat \varphi$. 
In particular in the case of $m=0$ (this is a physically reasonable case)
the bottom of the spectrum of $H_{\rm PF}$ lies in the bottom of the essential spectrum, and then it is not discrete. See  \cite{a18,gs11,hir19,sp04} as a review 
for ground states of models in quantum field theory.

The SRPF Hamiltonian is defined by $H_{\rm PF}$ with  kinetic term 
 $\frac{1}{2M}(\bp-\bA(\bx))^2$ replaced by 
a semi-relativistic version: 
 $$K_{\bA,M}=\sqrt{(\bp-\bA(\bx))^2+M^2}.$$ 
 It is of the form 
 \begin{align}
H_{M,m}=  K_{\bA,M} + \Hfm + V(\bx). \label{hamil000}
\end{align}
%where $K_{\bA,M}$ is formally given by  
%$$K_{\bA,M}=\sqrt{(\bp-\bA(\bx))^2+M^2}$$ is the kinetic energy of the 
%particle with the minimal coupling and $M\geq 0$ is the mass of the particle.
 It may also be further generalized to a model with $N$-charged particles for some $N>2$. 
 In the specific model studied here, we fix the number of the charged particle to one. 
 The SRPF Hamiltonian has two singularities: 
\begin{description}
\item[(zero photon mass)] $m=0$,
\item[(zero particle mass)] $M=0$. 
\end{description}
%In \cite{misp09, ms10, hisa10, hh16,kms11a, kms11b,km13a, km13b}
Hamiltonian $H_{0,m}$ is referred to as the SRPF Hamiltonian with a massless particle in this paper.   
The SRPF Hamiltonian with $(M,m)\not=(0,0)$ are studied so far. 
For example $H(0,m)$ with $m>0$ is studied in \cite{hh16} and $H_{M,0}$ with $M>0$ in the series of papers \cite{kms11a,kms11b,km13a,km13b,ms10}.
The analysis of SRPF Hamiltonian with $(M,m)=(0,0)$ however has been  left.  
The purpose of this paper is to investigate $H_{M,m}$  with 
$$(M,m)=(0,0).$$
In this case $H_{0,0}$ is denoted by 
 \begin{align}
H_{0,0}=  |\bp-{\bA(\bx)}|+ \Hf + V(\bx). \label{hamil0000}
\end{align}
The kinetic energy term  is of the form $|\bp-\bA(\bx)|$ 
which is a non-local operator and has a singularity in low energy.

\subsection{Technical improvement and the main result}
In \cite{hh16} it is shown that $H_{0,m}$ ($m>0$) has the normalized ground state 
$\Phi_m$ if external potential satisfies that $V(\bx)\to\infty$ as $|\bx|\to\infty$. Take a subsequence $m_j$ such that $\Phi_{m_j}$ weakly converges to 
some vector $\Phi$ as $m_j\to 0$ as $j\to\infty$. It is known that  if $\Phi\not=0$, then $\Phi$ is the ground state of $H_{0,0}$.  See \cite[Lemma 4.9]{ah97}. 

In order to establish $\Phi\not=0$, we improve methods developed by 
\cite{ge00,gll}.
We shall construct a compact operator $C$ such that 
$${\rm s}\!-\!\lim_{m_j\to0}C\Phi_{m_j}= C\Phi\not=0.$$
Let $j\in C_0^\infty([0,\infty))$ be a function such that $0\leq j(s)\leq 1$ and 
\begin{align}\label{jfunction}
  j(s) =
  \begin{cases}
    1 &  0 \leq s \leq 1, \\
    0 &  s \geq 2.
  \end{cases}
\end{align}
For $R>0$, 
let 
$\chi_1 = j(|\bx|/R)$,
$\chi_2 = j(|\bp|/R)$,
$\chi_3 = j(\Nf/R)$,
$\chi_4 = j(\Hf/R)$ and 
$\chi_5 = \Gamma(j(|i\nabla_\bk/R|))$.
Here $\Nf$ denotes the number operator and 
$\Gamma(j(|i\nabla_\bk/R|)) $ is the second quantization of $j(|i\nabla_\bk/R|)$. 
We can see that 
$C=\chi_1\chi_2\chi_3\chi_4\chi_5$ is compact and 
\begin{align}
\label{X}
 \sup_{j\in{\mathbb N}} \norm{(1-\chi_\ell)\Phi_{m_j}} = o(R^0), \quad \ell=1,\ldots,5
\end{align}
as $R\to\infty$. 
From this we shall show that 
$C\Phi_{m_j}\to C\Phi\not=0$ as $m_j\to\infty$, and we conclude that $H_{0,0}$ has the ground state. It is crucial to show cases of $\ell=3,5$ in \eqref{X};
\begin{align}
& \label{bd40}
 \lim_{R\to\infty}\sup_{j\in{\mathbb N}} \norm{(1-j(\Nf/R))\Phi_{m_j}} = 0,\\
&\label{bd50}
 \lim_{R\to\infty}\sup_{j\in{\mathbb N}} \norm{(1-\Gamma(j(|i\nabla_\bk/R|)) \Phi_{m_j}} = 0.
 \end{align}
We explain where the crucial part is and how to overcome the difficulties when studying $H_{0,0}$. 
The unperturbative  Hamiltonian associated with $H_{0,m}$ is given by 
$$H(0)=|\bp|+ \Hfm + V(\bx).$$
Hence the interaction of 
$H_{0,m}$ is the non-local operator of the form 
$$\hi=|\bp-{\bA(\bx)}|-|\bp|$$ and we have 
$$H_{0,m}=H(0) +\hi.$$
It is standard 
to apply the so-called pull-through formula to show \eqref{bd40}:
$$a(k)\Phi_m=(H_{0,m}-E_m+\omega(\bk))^{-1}[a(k),\hi]\Phi_m,$$
since $\|\Nf^{\frac{1}{2}}\Phi_m\|^2=\int\|a(k)\Phi_m\|^2 dk$. 
It is however hard to estimate $[a(k),\hi]$, since $\hi$ is singular ($M=0$) and non-local.
It is also unclear to specify the domains of both kinetic term 
$|\bp-{\bA(\bx)}|$ and commutator $[a(k),\hi]$. 
Moreover we cannot straightforwardly apply Pauli transformation 
\begin{align}
\label{PT}
U^{-1}(\bx) |\bp-\bA(\bx)| U(\bx)=|\bp+\bA(0)-\bA(\bx)|
\end{align}
 as was done for the Pauli-Fierz Hamiltonian $H_{\rm PF}$  in \cite{bfs3} to reduce the infrared divergence.  
It is indeed a bit hard to verify \eqref{PT} as an operator equality. 
We overcome these difficulties 
by combining functional integration (Proposition \ref{fk}), 
diamagnetic inequality (Lemma \ref{diamag}),  
 Hirokawa's trick  \eqref{xy243},   
 Hardy' inequality \eqref{HK} and Hardy-Kato's inequality \eqref{HK2}:
$$  \norm{|\bp|^{-\frac{1}{2}}|\Psi|}^2 \leq  \frac{\pi}{2} \norm{|\bx|^{\frac{1}{2}}\Psi}^2.$$
See e.g., \cite[Lemma 8.2]{ls10} and \cite{h77} for Hardy-Kato's inequality. 

Next to prove \eqref{bd50}
 we show that set $\{a(k)\Phi_m\}_{0<m<m_0}$ with some $m_0>0$ is equicontinuous in Theorem \ref{econt}. This is a Fock space-version of Kolmogorov-Riesz-Fr\'echet theorem, which proves that an equicontinuous set $D\subset L^p(\RR^d)$ is compact under some condition. 
 See e.g., \cite[Theorem 2.13 and Corollary 2.14]{hir19}.
 As far as we know this is new, and 
% we do not use the "photon derivative bound" developed in \cite{gll}. 
then we do not require extra regularity conditions  on $\hat\varphi$.

The main theorem is Theorem \ref{MainThm}, 
where it is assumed that the massive ground state $\Phi_m$ exist for each $m>0$ and the spatial decay of $\Phi_m$ is uniform in $m>0$.
This assumption is valid when $V(\bx)$ is a binding potential \cite{hh16}. 
In Theorem \ref{MainThm} the existence of the ground state of $H_{0,0}$ is shown. 

 \subsection{Previous results and organizations}
In e.g., \cite{misp09, ms10, hisa10, hh16,kms11a, kms11b,km13a, km13b}
the SRPF Hamiltonian is studied. 
The existence of the ground state for the SRPF Hamiltonian  is first proven by 
K\"onenberg, Matte and Stockmeyer \cite{kms11a} for $M>0$ and $m=0$.
As is proven in the non-relativistic Pauli-Fierz Hamiltonian, the bottom of the spectrum of $H_{M,0}$ coincides with that of its essential spectrum. 
%The Hamiltonian studied in \cite{kms11a} includes the spin of the particle,
 %and is not exactly the same as \eqref{hamil000}.
%A treatment of the two cases are different.
%An advantage of the spinless case is that various useful inequalitiescan be derived by the functional integration.
The case of $M=0$ but $m>0$ is investigated by Hidaka and Hiroshima \cite{hh16}, 
%Thus the spectral analysis for $M=0$ is more difficult rather than the case of $M>0$.
%The existence of ground state for $M=0$ was mentioned by Griesemer,
% Lieb and Loss \cite[P.560]{gll}.
where  
 $V(\bx)\to \infty (|\bx|\to\infty)$ is assumed and HVZ type theorem is shown.
In particular, for $m>0$, the ground state energy and the bottom of the essential spectrum of $H_m$ has a strictly positive gap,
and hence the ground state $\Phi_m$ of $H_{0,m}$ exists for each $m>0$.
The decaying potential $V(\bx)$ is not investigated in 
\cite{hh16}, the binding condition for the decaying potential is however proven in Hiroshima and Sasaki \cite{hisa10}.
Finally the uniqueness of the ground state is shown in \cite{hir14} for arbitrary $m\geq0$ and $M\geq0$ by a functional integration.   

This paper is organized as follows:

In Section \ref{sec:dmmr}, we give the  definition of the SRPF Hamiltonian  
and state the main theorem. 
%Sections \ref{sec:dpbo}--\ref{sec:slp} are the steps of the proof of the main theorem.
In Section \ref{sec:dpbo}, we discuss the bound and domain 
of $|\bp-\bA(\bx)|$. 
In Section \ref{sec:ptf}, we establish a singular and non-local pull-through formula. 
In Section \ref{sec:pmdb}, we estimate 
 $\|\Nf^{\frac{1}{2}} \Phi_m\|$ by the singular and non-local pull-through formula. 
In Section \ref{sec:slp}, we prove the spatial localization of $\Phi_m$ by 
showing that $\{a(k)\Phi_m\}_{0<m<m_0}$ is equicontinuous.  
%In order to prove it, we introduce a new general criterion to establish 
%the spatial localization which is a generalization of a method due to Gerard \cite{ge06}.
%This criterion will have many applications to other quantum field systems.
In Section \ref{sec:pmt}
 we prove the main theorem by compactness argument.

\section{Definition of SRPF model and main results} \label{sec:dmmr}
\subsection{Definition of SRPF model}
We define the Hamiltonian of SRPF model as 
 a self-adjoint operator acting in a Hilbert space over the complex field.
The operator consists of a particle part and a quantum field part. 
We firstly introduce the quantum field part.

The single photon Hilbert space is defined by 
$$W=L^2(\RR^3\times\{1,2\})$$
endowed with the inner product
\[
  \inner{f}{g} = \int  \overline{f(k)} g(k) dk,
\]
where $\int \ldots dk=\sum_{j=1,2}\int_{\RR^3} \ldots d\bk $ with $k=(\bk,j)\in\RR^3\times\{1,2\}$.
The boson Fock space over $W$ is given by
$\sF = \oplus_{n=0}^\infty \left[\stensor^n W\right]$,
where $ \stensor^n W$ denotes the symmetric tensor product of $W$ and
$\stensor^0 W=\CC$.
The inner product on $\sF$ is defined by $\inner{\Phi}{\Psi} =
\sum_{n=0}^\infty \inner{\Phi^{(n)}}{\Psi^{(n)}}_{\stensor^n W}$.
Thus $\Psi\in\sF$ can be identified with an  $\ell^2$-sequence $(\Psi^{(n)} )_{n=0}^\infty$ such that
$\sum_{n=0}^\infty \norm{\Psi^{(n)}}_{\stensor^n W}^2 <\infty$.
The Fock vacuum is the sequence defined by 
$$\Omega=(1,0,0,\ldots)\in\sF.$$
	Let $T$ be a densely defined closable operator in $W$. 
The second quantization of $T$ is a closed operator in $\sF$ defined by 
\begin{align*}
 d\Gamma(T) = \oplus_{n=0}^\infty \overline{T^{(n)}},
\end{align*}
where $T^{(n)}= {\sum_{j=1}^{n} \one\tensor\cdots \one\tensor
\stackrel{j\mathrm{th}}{T}\tensor \one\cdots \tensor \one}$ with $T^{(0)}=0$ and
$\overline{S}$ denotes the closure of closable operator $S$. 
 If $T$ is a non-negative self-adjoint operator in $W$, 
then $d\Gamma(T)$ turns to be also non-negative and self-adjoint.
We denote the spectrum (resp. point spectrum) of $T$ by 
$\spec(T)$ (resp. $\specP(T)$). 
The Fock vacuum $\Omega$ is an eigenvector of $d\Gamma(T)$ associated
with eigenvalue $0$, i.e., $d\Gamma(T)\Omega=0$. 
The number operator is defined by $\Nf=d\Gamma(\one)$.
Note that $\spec( \Nf ) = \NN\cup\{0\}$. 
Let $$\ome_m(\bk) = \sqrt{|\bk|^2+m^2}, \quad \bk\in\RR^3$$ be a dispersion relation and it
can be regarded as a multiplication operator in $W$. 
Here $m$ describes the mass of a single boson. 
Furthermore the free field Hamiltonian $\Hfm$ is given by the second quantization of $\omega_m$: 
\begin{align*}
  \Hfm = d\Gamma(\ome_m).
\end{align*}
We notice that $\Hfm$ is a non-negative self-adjoint operator in $\sF$, and 
the spectrum of $\Hfm$ is given by 
\begin{align*}
  \spec(\Hfm)=\{0\}\cup[m,\infty),\quad \specP(\Hfm)=\{0\}.
\end{align*}
For $m=0$, we write $\ome(\bk)=\ome_0(\bk)$ and $\Hf=d\Gamma(\ome)$.
The creation operator $a^\dag(f)$ smeared by $f\in W$ is given by
\begin{align*}
  (a^\dag(f)\Psi)^{(n)} = \sqrt{n}S_{n}(f\tensor \Psi^{(n-1)}),\;n\geq 1,
\end{align*}
and $(a^\dag(f)\Psi)^{(0)}=0$
with the domain:
\begin{align*}
  \dom(a^{\dagger}(f)) = \Big\{ \Psi \in \sF  \Mid
  \sum_{n=1}^{\infty} \norm{ \sqrt{n} S_{n}(f\tensor \Psi^{(n-1)}) }_{\stensor^n W}^{2}
  <\infty \Big\}.
\end{align*}
Here $S_{n}$ is the symmetrization operator on $\tensor^{n} W$.
The annihilation operator smeared by $f=f(k)=f(\bk,j)\in W$ is defined by the adjoint of
$a^\dag(\bar{f})$: $a(f)=(a^\dag(\bar{f}))^{*}$.
Both $a(f)$ and $a^\dag(f)$ are linear in $f$,
and satisfy canonical commutation relations: 
\begin{align*}
  [a(f),a^\dag(g)]= \inner{\bar f}{g}_W, \qquad [a(f),a(g)]=0=[a^\dag(f),a^\dag(g)].  
\end{align*}
We informally write 
$a^\sharp(f) = \int a^\sharp(k) f(k) dk=
\sum_{j=1,2}\int_{\RR^3}a^\sharp(\bk,j)f(\bk,j)d\bk$ 
for $a^\sharp(f)$.
Let us introduce the finite particle subspace $\Ffin$ by 
\begin{align*}
\Ffin={\rm L.H.}\{\Omega, a^\dag (h_1)\cdots a^\dag(h_n) 
\Omega \mid h_j \in C_0^\infty (\RR^3\times\{1,2\}), j=1,\ldots,n, n\geq 1\},
\end{align*}
where $C_0^\infty(\RR^3\times\{1,2\}) = C_0^\infty (\RR^3) \oplus C_0^\infty (\RR^3)$.
Note that $\Ffin$ is dense in $\sF$.
Next we shall define the quantized radiation field $\bA(\bx)$ for each $\bx\in\RR^3$. 
Let $\be(\bk,j)$ be polarization vectors, which is defined by
\begin{align*}
 \be(\bk,1) = \frac{(k_2,k_1,0)}{\sqrt{k_1^2+k_2^2}}, 
\qquad 
 \be(\bk,2) = \frac{\bk}{|\bk|}\times \be(\bk,1).
\end{align*}
Note that $\be(\bk,j), j=1,2$ satisfy
\begin{align*}
 \bk\cdot\be(\bk,j)=0, \qquad \be(\bk,j)\cdot\be(\bk,j')=\delta_{jj'},
 \qquad j,j'=1,2.
\end{align*}
We write $\be(\cdot)=(e_1(\cdot),e_2(\cdot),e_3(\cdot))$.
Note that $e_\mu(\cdot,j)\in C^\infty(\RR^3\setminus L_{12})$, where 
$$L_{12}=\{\bk=(k_1,k_2,k_3)\in\RR^3\mid k_1=k_2\}.$$ 
The quantized radiation field  $\bA(\bx)=(A_1(\bx),A_2(\bx),A_3(\bx))$ is defined by
\begin{align*}
  A_\mu(\bx) 
 = \frac{1}{\sqrt{2}} \sum_{j=1,2} \int_{\RR^3} e_\mu(\bk,j) 
   (a^\dag(\bk,j) \phiome(\bk) e^{-i\bk\cdot\bx} + a(\bk,j) \phiome(-\bk) e^{+i\bk\cdot\bx}) d\bk ,
\end{align*}
where the function $\phiome$ has the form
\begin{align*}
  \phiome(\bk) = \frac{\phih(\bk)}{\sqrt{\ome(\bk)}},
\end{align*}
and $\phih(\bk)$ is called an ultraviolet cutoff function.
Let us introduce assumptions on $\phih$.
\begin{enumerate}
  \item[\textbf{(A1)}] 
    $\phih(\bk)=\overline{\phih(-\bk)}$ and $\ome^{-\frac{1}{2}}\phih \in L^2(\RR^3)$.
  \item[\textbf{(A2)}] 
    $\ome^{-1}\phih \in L^2(\RR^3)$ and $\ome^{\frac{3}{2}}\phih \in L^2(\RR^3)$.
\end{enumerate}
By assumption (A1), $A_\mu(\bx)$ is essentially self-adjoint on $\Ffin$ 
for each $\bx \in \RR^3$. 
We denote the closure of $A_\mu(\bx)$ by the same symbol.
The assumption (A2) will be used for the self-adjointness of the total Hamiltonian.

Next we  explain the particle part.
The Hilbert space for the particle is $L^2(\RR_\bx^3)=L^2(\RR^3,d\bx)$, where
$\bx=(x_1,x_2,x_3)\in\RR^3$ denotes the position of the particle.
Let $\bp=(p_1,p_2,p_3)=-i(\del_{x_1},\del_{x_2},\del_{x_3})$ be the momentum operator
of the particle. 
The particle Hamiltonian under consideration is a relativistic Schr\"odinger
operator given by 
\begin{align*}
  H_\mathrm{p} = \sqrt{|\bp|^2+M^2}+V(\bx) = \sqrt{-\Delta +M^2}+V(\bx),
\end{align*}
%which acts on $L^2(\RR^3_\bx)$, 
where
$M\geq 0$ denotes the mass of the particle and $V:\RR_\bx^3\to\RR$ is an external potential. 

The Hilbert space for SRPF model is defined by
\begin{align*}
 \sH = L^2(\RR^3_\bx) \tensor \sF.
\end{align*}
We  use the identification until confusions may arise:
\begin{align*}
 \sH \cong L^2(\RR_\bx^3; \sF) \cong \int^\oplus_{\RR^3} \sF d\bx.
\end{align*}
Under this identification, we can define the constant fiber direct integral 
$\int_{\RR^3}^\oplus A_\mu(\bx) d\bx$, which is also denoted by $A_\mu(\bx)$ for simplicity.
Then $A_\mu(\bx),\mu=1,2,3$ are self-adjoint operators in $\sH$.
The interaction between the particle and quantized radiation field is described
by the minimal coupling, i.e., the interacting Hamiltonian is
obtained by replacing $\bp$ by $\bp-\bA(\bx)$.
Thus the total Hamiltonian of SRPF model with particle mass $M$ and 
photon mass $m$ is formally defined by
\begin{align*}
  H_{M,m} = \sqrt{(\bp\tensor\one-\bA(\bx))^2+M^2} + \one\tensor \Hfm+ V(\bx)\tensor\one.
\end{align*}
We do not write tensor notation $\tensor$ for notational convenience in what follows.
Thus $H_{M,m}$ can be simply written as
\begin{align*}
  H_{M,m} =\sqrt{(\bp-\bA(\bx))^2+M^2} + V(\bx) + \Hfm.  
\end{align*}
Note that the definition of $H_{M,m}$ is currently unclear, and we have to specify
the definition of the square root appearing in $H_{M,m}$ and conditions for $V(\bx)$.
We use the notation that $C^\infty(T)=\cap_{n=1}^\infty \dom(T^n)$ for operator $T$.
By assumption (A2), the non-relativistic kinetic energy
\begin{align*}
 T_\bA = (\bp-\bA(\bx))^2
\end{align*}
is well defined on $\dom(|\bp|^2)\cap C^\infty( \Nf )$, and 
the next proposition has been  established.
%%%%%%%%%%%%%%%%%%%%%%%%%%%%%%%%%%%%%%%%%%%%%%%%%%%%%%%%%%%%%%%%
\begin{prop}[{\cite[Proposition 3.4]{hir14}}] \label{ess1}
 Assume (A1) and (A2). 
Then $T_\bA$ is essentially self-adjoint on $\dom(|\bp|^2) \cap C^\infty(\Nf)$.
\end{prop}
%%%%%%%%%%%%%%%%%%%%%%%%%%%%%%%%%%%%%%%%%%%%%%%%%%%%%%%%%%%%%%%%
We set
\begin{align*}
 \Hfin = \Czinf \hat\tensor \Ffin,
\end{align*}
where $\hat\tensor$ denotes the algebraic tensor product.
Proposition \ref{ess1} can be extended:
% to a general The next proposition is more useful than the previous result.
%%%%%%%%%%%%%%%%%%%%%%%%%%%%%%%%%%%%%%%%%%%%%%%%%%%%%%%%%%%%%%%%
\begin{prop}
 Assume (A1) and (A2). 
Then $T_\bA$ is essentially self-adjoint on $\Hfin$.
\end{prop}
%%%%%%%%%%%%%%%%%%%%%%%%%%%%%%%%%%%%%%%%%%%%%%%%%%%%%%%%%%%%%%%%
\begin{proof}
Set $\cD_1 = \dom(|\bp|^2)\cap C^\infty(\Nf)$. 
Then, by Proposition \ref{ess1}, $\overline{T_\bA\lceil \cD_1}$ is 
self-adjoint.
We use the fact that $\Hfin$ is a core for $|\bp|^2+\Nf$.
Let $\Psi \in \cD_1$.
Then $\Psi \in\dom(|\bp|^2+\Nf)$, and hence there exists a sequence
$\{\Psi_n\}_n \subset \Hfin$ such that
$\Psi_n\to \Psi$ and $(|\bp|^2+\Nf) \Psi_n \to (|\bp|^2+\Nf) \Psi$
as $n\to\infty$. On the other hand, for $\Phi\in \Hfin$, we have
\begin{align}
 \norm{T_\bA \Phi}
 = \norm{(|\bp|^2-2\bA(\bx)\cdot\bp+\bA(\bx)^2) \Phi}
 \leq a \norm{(|\bp|^2+\Nf)\Phi} + b\norm{\Phi}  \label{yy112}
\end{align}
for some $a,b>0$.
From \eqref{yy112}, we know that $\{T_\bA\Psi_n\}_n$ is a convergent sequence.
Therefore $\Psi\in \dom(\overline{T_\bA\lceil \Hfin})$, which
means that $T_\bA \lceil \cD_1 \subset \overline{T_\bA\lceil\Hfin}$.
Since the self-adjoint extension is unique, we have 
$\overline{T_\bA\lceil\Hfin} = \overline{T_\bA \lceil \cD_1}$ which is self-adjoint.
\end{proof}
%%%%%%%%%%%%%%%%%%%%%%%%%%%%%%%%%%%%%%%%%%%%%%%%%%%%%%%%%%%%%%%%
We denote the closure of $T_\bA$ by the same symbol and 
the relativistic kinetic energy $\sqrt{(\bp-\bA(\bx))^2+M^2}$ 
is defined through the spectral measure of $T_\bA$, i.e., 
$$\sqrt{(\bp-\bA(\bx))^2+M^2}=
\sqrt{T_\bA+M^2}.$$
%%%%%%%%%%%%%%%%%%%%%%%%%%%%%%%%%%%%%%%%%%%%%%%%%%%%%%%%%%%%%%%%
\begin{df}[SRPF Hamiltonian]
SRPF Hamiltonian is defined by 
\begin{align}
H_{M,m}=
\sqrt{T_\bA+M^2}+V+\Hfm.
\end{align}
\end{df}
We write 
$\sqrt{(\bp-\bA(\bx))^2+M^2}$ for $\sqrt{T_\bA+M^2}$, and 
$|\bp-\bA(\bx)|$ for $\sqrt{T_\bA}$ in what follows.  
 We set 
\begin{align*}
&H_m=H_{0,m}=|\bp-\bA(\bx)| + V(\bx) + \Hfm,\\
&H=H_{0,0}=|\bp-\bA(\bx)| + V(\bx) + \Hf.
\end{align*}
The main object in this paper is to study the spectrum of  $H$, and 
in particular  we study the existence of the ground state of $H$.
%The massive Hamiltonians, i.e., $H_{M,m}$ with $(M,m)\neq (0,0)$, will be used as a regularization of $H$.

\subsection{The main results}
We define two classes of external potentials.
\begin{df}
\begin{enumerate}
\item [\rm(1)]
$V\in 
 V_\mathrm{rel} $ if and only if 
 $\dom(H_{\rm p})\subset \dom(V)$ and there exist $0\leq a<1$ and $0\leq b$ such that 
 $\|Vf\|\leq a\|H_{\rm p}f\|+b\|f\|$ for any $f\in \dom(H_{\rm p})$.
 \item [\rm (2)] 
 $V\in  V_\mathrm{conf} $ if and only if 
 $\lim_{|x|\to\infty}V(x)=\infty$, $\dom(V)\subset \dom(|\bx|)$, 
and $V \in C^2(\RR^3)$ with 
$ \del_\mu V, \del_\mu^2 V 
     \in L^\infty(\RR^3)$ for $\mu=1,2,3$.
     \end{enumerate}
\end{df}
%Note that $V\in V_\mathrm{conf}$ satisfies $\dom(V)\subset \dom(|\bx|)$.
Examples of $V_\mathrm{rel}$ and $V_\mathrm{conf}$ are
$-Z/|\bx|\in V_\mathrm{rel}$ and $\braket{\bx}=\sqrt{1+|\bx|^2}\in V_\mathrm{conf}$.
%%%%%%%%%%%%%%%%%%%%%%%%%%%%%%%%%%%%%%%%%%%%%%%%%%%%%%%%%%%%%%%%
\begin{prop}[{\cite[Theorem 1.9]{hh15}}]
Assume (A1) and (A2). Suppose that $V\in V_\mathrm{conf} \cup V_\mathrm{rel}$. 
Then, for any $m\geq0$ and $M\geq 0$, $H_{M,m}$ is self-adjoint on $\dom(|\bp|)\cap \dom(V)\cap \dom(\Hfm)$
and essentially self-adjoint on $\Hfin$.
\end{prop}
%%%%%%%%%%%%%%%%%%%%%%%%%%%%%%%%%%%%%%%%%%%%%%%%%%%%%%%%%%%%%%%%

If $T$ is self-adjoint and bounded from below, then
an eigenvector $f$ such that $Tf=Ef$ with $E=\inf\spec(T)$ is called a ground state of $T$. 
%We will construct the ground state of $H$ from 
The existence of the ground state of the 
massive Hamiltonian $H_m$ 
has been established:
%%%%%%%%%%%%%%%%%%%%%%%%%%%%%%%%%%%%%%%%%%%%%%%%%%%%%%%%%%%%%%%%
\begin{prop}[{\cite[Theorem 2.8]{hh16},\cite[Theorem 5.12 (2)]{hir14}}]
\label{massive result}
 Assume (A1) and (A2). Suppose that $V\in V_\mathrm{conf}$.
Then $H_m$ has a  ground state $\Phi_m$ for each $m>0$, and 
there exist $C$ and $c$ such that 
%\begin{align}
%  \sup_{m>0}\norm{\Phi_m(\bx)}_\sF \leq C \braket{\bx}^{-3-1}, \qquad \bx\in\RR^3, \label{decay p}
%\end{align}
%where $\braket{\bx}=(1+|\bx|^2)^{\frac{1}{2}}$.
\begin{align}
  \sup_{m>0}\norm{\Phi_m(\bx)}_\sF \leq 
  C e^{-c|\bx|}, \qquad \bx\in\RR^3. \label{decay p}
\end{align}
%In particular, $\Phi_m \in \dom(|\bx|^2)$ and 
%\begin{align}\sup_{m>0} \norm{\braket{\bx}^2\Phi_m} < \infty.   \label{decay p2}\end{align}
\end{prop}
%%%%%%%%%%%%%%%%%%%%%%%%%%%%%%%%%%%%%%%%%%%%%%%%%%%%%%%%%%%%%%%%
\begin{rem}
In Proposition \ref{massive result} it is assumed that $V$ is a confining potential.  
In \cite[Theorem~5.12 (1)]{hir14} however 
a spatial decay of bound states of $H_m$ with a decaying potential are shown for  $m\geq0$. Let $H_m\Psi=E_m\Psi$. 
Suppose that $V$ is negative and 
$\lim_{|\bx|\to\infty}E_m-V(x)<0$. 
Then $$\|\Psi(x)\|_{\sF}\leq \left\{\begin{array}{ll}
C\braket{\bx}^{-3-1}&m=0,\\
C_m e^{-c_m|\bx|}&m>0
\end{array}
\right.$$
with some constants $c_m,C_m$ and $C$. 
\end{rem}
One dominant method to prove the existence of the ground state of $H$ 
is to show that the weak limit of $\Phi_m$ as $m\to0$ is a non-zero vector $\Phi$. 
In Proposition \ref{massive result}
under some condition on $V$ and cutoff it is shown that $H_m$ has  
the ground state $\Phi_m$ for each $m>0$.
%The ground state of the massless Hamiltonian $H$ will be constructed by a limit of 
%$\Phi_m$, where the bound \eqref{decay p2} plays an important role.
%One might expect that Theorem \ref{massive result} can be extended for a wider class of potentials.
%We separate the problem of its extension from the proof of the existence of a massless ground state, which clarify the structure of the massless limit.
Thus  in this paper, we investigate  the  limit of $\Phi_m$ under the following general conditions:
%%%%%%%%%%%%%%%%%%%%%%%%%%%%%%%%%%%%%%%%%%%%%%%%%%%%%%%%%%%%%%%%
\begin{enumerate} 
 \item[\textbf{(A3)}] For any $m>0$, $H_m$ has a normalized ground state $\Phi_m$.
 \item[\textbf{(A4)}] There exists $m_0>0$ such that
     $\sup_{0<m<m_0} \norm{\braket{\bx}^2 \Phi_m} < \infty$. 
\end{enumerate}
%%%%%%%%%%%%%%%%%%%%%%%%%%%%%%%%%%%%%%%%%%%%%%%%%%%%%%%%%%%%%%%%

The main result in this paper is the following:

%%%%%%%%%%%%%%%%%%%%%%%%%%%%%%%%%%%%%%%%%%%%%%%%%%%%%%%%%%%%%%%%
\begin{theorem}{\label{MainThm}}
  Assume (A1)--(A4) and  $V\in V_\mathrm{conf} \cup V_\mathrm{rel}$.
 Then $H$ has the ground state.
\end{theorem}
%%%%%%%%%%%%%%%%%%%%%%%%%%%%%%%%%%%%%%%%%%%%%%%%%%%%%%%%%%%%%%%%

%%%%%%%%%%%%%%%%%%%%%%%%%%%%%%%%%%%%%%%%%%%%%%%%%%%%%%%%%%%%%%%%
\section{Domains and bounds of $|\bp-\bA(\bx)|$}\label{sec:dpbo}
In this section, we discuss domains and bounds of operators related to 
$(\bp-\bA(\bx))^2$. 
In the spectral analysis of $H$, we need to compute and estimate commutators related to 
$\sqrt{(\bp-\bA(\bx))^2+M^2}$.
Since $\sqrt{(\bp-\bA(\bx))^2+M^2}$ is non-local, 
it is not apparent that $\Nf^{\frac{1}{2}}\sqrt{(\bp-\bA(\bx))^2+M^2}$ is  well defined on a dense domain.

Let $\Omega(x)=\pi^{-\frac{1}{4}} e^{-\frac{1}{2}x}$. Intuitively 
in the case of one mode annihilation operator and creation operator $a=(x+d/dx)/\sqrt 2$ and $a^\dag=(x-d/dx)/\sqrt 2$ in $L^2(\RR)$, we have 
$$|a+a^\dag|\Omega = \sqrt 2 \pi^{-\frac{1}{4}} |x|e^{-\frac{1}{2}x}$$ which is not twice differentiable, because of 
the singularity at 
$x=0$.
Namely $$|a+a^\dag|\Omega \notin \dom(a^\dag a) = 
\dom(-\frac{1}{2}\frac{d^2}{dx^2}+\frac{1}{2}x^2-\frac{1}{2}).$$
From this observation  
$|\bp-\bA(\bx)|\Psi \in \dom(\Nf )$ may not be expected for $\Psi\in\Hfin$.
Since we can see however that 
$$|a+a^\dag|\Omega \in  \dom((a^\dag a)^{\frac{1}{2}})=\dom(\frac{d}{dx})\cap\dom(x),$$
we may expect that $|\bp-\bA(\bx)| \Psi \in \dom(\Nf ^{\frac{1}{2}})$ for $\Psi\in\Hfin$.  
We can indeed show the proposition below:
%%%%%%%%%%%%%%%%%%%%%%%%%%%%%%%%%%%%%%%%%%%%%%%%%%%%%%%%%%%%%%%
\begin{prop}{\label{dom0}}
 Suppose (A1) and (A2).
 Then $|\bp-\bA(\bx)| \Psi \in \dom(\Nf^{\frac{1}{2}})$ for any $\Psi\in\Hfin$.
\end{prop}
%%%%%%%%%%%%%%%%%%%%%%%%%%%%%%%%%%%%%%%%%%%%%%%%%%%%%%%%%%%%%%%%%
The proof will be given later in this section.
The next lemma is a basic fact about the domains related to $T_\bA$ and $\Nf$.
%%%%%%%%%%%%%%%%%%%%%%%%%%%%%%%%%%%%%%%%%%%%%%%%%%%%%%%%%%%%%%%%
\begin{lem}{\label{dom1}}
  Assume (A1) and (A2). If $\Psi\in\Hfin$, then $\Psi\in\dom(T_\bA^2)$
  and $T_\bA^2\Psi \in C^\infty(\Nf)$.
\end{lem}
\begin{proof}
Note that $\Hfin\subset \dom(|\bp|^2)\cap C^\infty(\Nf) \subset \dom(T_\bA)$.
By the properties of polarization vectors, we know $\bA(\bx)\cdot\bp=\bp\cdot\bA(\bx)$, so
$T_\bA\Psi =(|\bp|^2-2\bA(\bx)\cdot\bp+\bA(\bx)^2)\Psi$ for $\Psi\in\Hfin$.
By (A2), we have $|\bk|^2\phiome\in L^2(\RR^3)$, which means that  
$A_\mu(\bx) \Phi \in \dom(|\bp|^2)$ if $\Phi\in\dom(|\bp|^2)\cap\dom(\Nf^{\frac{1}{2}})$.
Hence $|\bp|^2\Psi, \bA(\bx)\cdot\bp\Psi, \bA(\bx)^2\Psi \in \dom(|\bp|^2)$.
Clearly, each vectors have finite photon number.
Thus $T_\bA\Psi\in \dom(|\bp|^2)\cap C^\infty(\Nf)\subset\dom(T_\bA)$,
and $T_\bA\Psi \in \dom(T_\bA)$. It is clear that $T_\bA^2\Psi \in C^\infty(\Nf)$.
\end{proof}
%%%%%%%%%%%%%%%%%%%%%%%%%%%%%%%%%%%%%%%%%%%%%%%%%%%%%%%%%%%%%%%%

In order to prove Proposition \ref{dom0}, we need some inequalities
derived by the functional integral representation.
We consider the probabilistic representation.
Let $(B_t)_{t\geq 0}$ be the three dimensional Brownian motion on a 
probability space $(\cW, B(\cW), P^\bx)$.
Here $P^\bx$ is the Wiener measure starting from $\bx \in \RR^3$.
Then we can consider the partial isometry
\begin{align}
 L^2(\RR^3,d\bx ) & \to  \int_{\RR^3}^\oplus L^2(\cW,dP^\bx) d\bx,   \label{BW} \\
 f(\bx) & \mapsto  f( B_0(w)), \qquad (\bx,w)  \in \RR^3\times \cW.
\end{align}
Since $B_0(w)=\bx$ a.s., the above identification is trivial.
However, the semigroup for the free particle can be described as 
\begin{align*}
  ( e^{-\frac{t}{2}|\bp|^2} f )(\bx) \mapsto f(\bx + B_t(w)), \qquad (\bx,w)\in \RR^3\times \cW.
\end{align*}
The expectation with respect to $P^\bx$ is simply denoted by $\EE^\bx [\ldots]$.
In the following we use this embedding \eqref{BW} as an identification, and we simply use 
$L^2(\RR^3\times \cW)$ to denote $\int_{\RR^3}^\oplus L^2(\cW,dP^\bx) d\bx $.
Next we introduce a probabilistic description for the field.
Let $\cA(F)$ be the Gaussian random process indexed by $F\in \oplus^3 L^2(\RR^3)$
on a probability space $(Q, \Sigma,\mu)$ such that $\EE_\mu[\cA(F)]=0$
and the covariance is given by 
\begin{align*}
\EE_\mu[\cA(F)\cA(G)]
 = \half \sum_{\mu,\nu=1}^3\inner{\hat{F}_\mu}{d_{\mu\nu}\hat{G}_\nu},
\end{align*}
where $d_{\mu\nu}=\delta_{\mu\nu} - k_\mu k_\nu / |\bk|^2$ and $\hat{F}_\mu$
denotes the Fourier transform of $F_\mu$.
The unitary equivalence between $L^2(Q)$ and $\sF$ is established,
and under this equivalence it follows that for 
$F=F_1\oplus F_2\oplus F_3 \in \oplus^3 L^2(\RR^3)$,
\begin{align}
 \cA(F) \cong A(F)=
 \frac{1}{\sqrt{2}} \sum_{\mu=1}^3 \sum_{j=1,2}
 \int_{\RR^3} e_\mu(\bk,j) 
 \big (a^\dag (\bk,j) \hat{F}_\mu(\bk) + a(\bk,j)\hat{F}_\mu(-\bk) \big) d\bk.  \label{AF}
\end{align}
Namely, each Segal's field operator can be considered as a Gaussian random process.
In the following, we use the identifications $L^2(\RR^3,d\bx ) \to L^2(\RR^3 \times \cW)$
 and $\sF \cong L^2(Q)$.
%%%%%%%%%%%%%%%%%%%%%%%%%%%%%%%%%%%%%%%%%%%%%%%%%%%%%%%%%%%%%%%%
\begin{prop}[\cite{hir00}]{\label{fk}}
 The Feynman-Kac formula  of $e^{-\frac{t}{2} T_\bA}$ is given by
\begin{align*}
  \inner{\Phi}{e^{-\frac{t}{2} T_\bA}\Psi}
  = \int_{\RR^3} \EE^\bx \left[\inner{\Phi(B_0)}{e^{-i\cA(K)}\Psi(B_t)}_{L^2(Q)} \right] d\bx,
  \qquad \Psi,\Phi \in \sH.
\end{align*}
Here
\begin{align}
  K(\cdot) = \oplus_{\mu=1}^3 \int_0^t \til\varphi(\cdot - B_s)  dB_s^\mu   \label{defK}
\end{align}
with $\til\varphi = (\phiome)\check{} = (\phih/\sqrt{\ome})\check{}$.
\end{prop}
%%%%%%%%%%%%%%%%%%%%%%%%%%%%%%%%%%%%%%%%%%%%%%%%%%%%%%%%%%%%%%%%
Let $\cN$ be the number operator in $L^2(Q)$.
For $F\in \oplus^3 L^2(\RR^3)$, the conjugate momentum of $\cA(F)$ is denoted by $\Pi(F)$,
namely, $ \Pi(F) = i[\cN, \cA(F)]$ and the corresponding field operator is
\begin{align*}
 \pi(F) = \frac{i}{\sqrt{2}} \sum_{\mu=1}^3 \sum_{j=1,2}
 \int_{\RR^3} e_\mu(\bk,j) (a^\dag (\bk,j) \hat{F}_\mu(\bk) - a(\bk,j)
 \hat{F}_\mu(-\bk)) d\bk.
\end{align*}
Then the identity 
\begin{align}
 \cN e^{-i\cA(K)} = e^{-i\cA(K)}( \cN -\Pi(K) -\xi_K)   \label{throughN}
\end{align}
holds, where $\xi_K$ is a stochastic process defined by
\begin{align*}
  \xi_K 
= \half \sum_{\mu,\nu=1}^3 
   \inner{\hat{K}_\mu}{ d_{\mu\nu} \hat{K}_\mu}_{L^2(\RR^3)}.
\end{align*}
Note that $\hat{K}_\mu = \int_0^t \phi_\ome(\bk) e^{-i\bk \cdot B_s} d B_s^\mu$ is 
an 
$L^2(\RR_\bk^3)$-valued stochastic integral, and hence $\pi(K)$ is an operator-valued 
stochastic integral in $L^2(\RR^3\times\cW)\tensor \sF$.
Let 
$$P_\mu = p_\mu \tensor \one + \one \tensor P_{{\rm f}\mu},\quad \mu=1,2,3$$ 
be the total momentum, where $P_{{\rm f}\mu}=d\Gamma(k_\mu)$ is the field momentum. 
The corresponding filed momentum in $L(Q)$ is denoted by $\cP_{{\rm f}\mu}$. 
The commutation relation between $\cP_{{\rm f}\nu}$ and $e^{-i\cA(K)}$ is given by
\begin{align*}
  \cP_{{\rm f}\nu} e^{-i\cA(K)} = e^{-i\cA(K)} ( \cP_{{\rm f}\nu} - \cA(\del_\nu K) ),  
\end{align*}
where the last term is obtained from $\cA(\del_\nu K) = i[\cP_{{\rm f}\nu}, \cA(K)]$, and 
the corresponding field operator is
\begin{align*}
 \cA(\del_\nu K) \cong
 \frac{1}{\sqrt{2}} \sum_{\mu=1}^3 \sum_{j=1,2}
 \int_{\RR^3} e_\mu(\bk,j) (a^\dag (\bk,j) (i k_\nu \hat{F}_\mu)(\bk) 
 + a(\bk,j) (i k_\nu \hat{F}_\mu)(-\bk)) d\bk.
\end{align*}
Note that $\del_\nu$ in the above expression means the derivative for the photon coordinate.

Let $U_\sF:\sF\to L^2(Q)$ be the unitary operator implementing the identification $\sF\cong L^2(Q)$.
Then $(\one\tensor U_\sF) \Psi~(\Psi \in \cH)$ is a function in $L^2(\RR^3_\bx\times Q)$
and the absolute value of $\Psi$ is defined under this identification.
The following is a variation of diamagnetic inequalities.
%%%%%%%%%%%%%%%%%%%%%%%%%%%%%%%%%%%%%%%%%%%%%%%%%%%%%%%%%%%%%%%%%
\begin{lem}{\label{diamag}}
Assume (A1) and (A2). Then
\begin{enumerate}
\item[\rm (1)] For any $\Psi\in \sH$,
  \begin{align*}
    \norm{(T_{\bA}+s)^{-\frac{1}{2}}\Psi} \leq \norm{(|\bp|^2+s)^{-\frac{1}{2}}|\Psi|}, \qquad s>0.
  \end{align*}
\item[\rm (2)] If $\Psi\in \dom(|\bx|)$, then $\Psi \in \dom(T_\bA^{-\frac{1}{2}})$ and
  it holds that \begin{align*}
    \norm{T_\bA^{-\frac{1}{2}}\Psi} \leq 2 \norm{|\bx| \Psi}.
  \end{align*}
 \item[\rm (3)] Let $\varrho=\varrho(\bx)$ be a function of $\bx$ and $s>0$.
Suppose that $\norm{\varrho(|\bp|^2+s)^{-1}|\Psi|}<\infty$. Then 
$(T_\bA+s)^{-1}\Psi \in \dom(\varrho)$ and it holds that 
  \begin{align}
    \norm{\varrho(T_\bA+s)^{-1}\Psi}
    \leq \norm{\varrho(|\bp|^2+s)^{-1}|\Psi|}.  \label{diamag2}
  \end{align}
  \end{enumerate}
\end{lem}
%%%%%%%%%%%%%%%%%%%%%%%%%%%%%%%%%%%%%%%%%%%%%%%%%%%%%%%%%%%%%%%%
\begin{proof}
 By Proposition \ref{fk}, we have
\begin{align*}
  \norm{(T_\bA+s)^{-\frac{1}{2}}\Psi}^2
& = \frac{1}{2} \int_0^\infty  e^{-\frac{ts}{2}} \inner{\Psi}{e^{-\frac{t}{2} T_\bA}\Psi} dt\\
& = \frac{1}{2} \int_0^\infty e^{-\frac{ts}{2}} dt \int_{\RR^3}
    \EE^\bx \big[ \inner{\Psi(B_0)}{e^{-i \cA(K)}\Psi(B_t)}_{L^2(Q)}  \big] d\bx  \\
& \leq \frac{1}{2} \int_0^\infty e^{-\frac{ts}{2}} dt \int_{\RR^3} 
    \EE^\bx \big[ \inner{|\Psi(B_0)|}{|\Psi(B_t)|}_{L^2(Q)}  \big] d\bx  \\
& = \frac{1}{2} \int_0^\infty e^{-\frac{ts}{2}} \biginner{|\Psi|}{e^{-\frac{t}{2}|\bp|^2}|\Psi|} dt
 = \norm{(|\bp|^2+s)^{-\frac{1}{2}}|\Psi|}^2.
\end{align*}
Thus (1) follows.
Next we assume that $\Psi \in \dom(|\bx|)$. Clearly $|\Psi| \in \dom(|\bx|)$ and 
by Hardy's inequality, we have $|\Psi| \in \dom(|\bp|^{-1})$ and 
\begin{align}\label{HK}
  \norm{|\bp|^{-1}|\Psi|} \leq 2\norm{|\bx| |\Psi|} = 2\norm{|\bx|\Psi}.
\end{align}
By (1) and the monotone convergence theorem, we have $\Psi \in \dom(T_\bA^{-\frac{1}{2}})$ and 
\begin{align*}
 \norm{T_\bA^{-\frac{1}{2}}\Psi} 
 = \lim_{s\to +0} \norm{(T_\bA+s)^{-\frac{1}{2}}\Psi}
 \leq \lim_{s\to +0} \norm{(|\bp|^2+s)^{-\frac{1}{2}}|\Psi|}
\leq 2 \norm{|\bx| \Psi},
\end{align*}
which proves (2). Next we prove (3). %Write $\varrho= \varrho(\bx)$.
By the Feynman-Kac formula (Proposition \ref{fk}), we have
\begin{align*}
& \norm{\varrho(\bx)(T_\bA+s)^{-1}\Psi}  = \sup_{\Phi\in \dom(\varrho^\ast), \norm{\Phi}=1} \big|\inner{\varrho^*\Phi}{(T_\bA+s)^{-1}\Psi} \big| \\
& = \sup_{\Phi\in \dom(\varrho^\ast), \norm{\Phi}=1} \bigg| \frac{1}{2} \int_0^\infty e^{-\frac{ts}{2}} dt
    \int_{\RR^3} \EE^\bx \big[ \inner{(\varrho^*\Phi)(B_0)}{e^{-i \cA(K)}\Psi(B_t)}_{L^2(Q)} \big] d\bx  \bigg| \\
& \leq  \sup_{\Phi\in \dom(\varrho^\ast), \norm{\Phi}=1}\frac{1}{2} \int_0^\infty e^{-\frac{ts}{2}} dt
  \int_{\RR^3} \EE^\bx \big[ \inner{|(\varrho^*\Phi)(B_0)|}{|\Psi(B_t)|}_{L^2(Q)} \big] d\bx \\
& =  \sup_{\Phi\in \dom(\varrho^\ast), \norm{\Phi}=1} \frac{1}{2} \int_0^\infty e^{-\frac{ts}{2}} 
    \biginner{|\varrho| |\Phi|}{e^{-\frac{t}{2}|\bp|^2}|\Psi|} dt \\
& = \sup_{\Phi\in \dom(\varrho^\ast), \norm{\Phi}=1}  \inner{|\varrho||\Phi|}{(|\bp|^2+s)^{-1}|\Psi|} \leq  \norm{|\varrho|(|\bp|^2+s)^{-1}|\Psi|},
\end{align*}
which proves (3).
\end{proof}
%%%%%%%%%%%%%%%%%%%%%%%%%%%%%%%%%%%%%%%%%%%%%%%%%%%%%%%%%%%%%%%%

%%%%%%%%%%%%%%%%%%%%%%%%%%%%%%%%%%%%%%%%%%%%%%%%%%%%%%%%%%%%%%%%
\begin{lem}{\label{thB}}
Assume (A1) and (A2).
Let $K$ be $\oplus^3 L^2(\RR^3)$-valued stochastic integral given by \eqref{defK}.
Suppose that $ \Phi\in \dom(\Nf^k)$. 
Then, for $k\in \NN$, there exists a polynomial $P_k = P_k(\tau)$
 of degree $k$ such that
\begin{equation}
  \norm{(\Nf-\pi(K)-\xi_K)^k\Phi}_\sF 
  \leq P_k(|\xi_K|) \norm{(\Nf+\one)^k \Phi}_\sF. \label{polyb}
\end{equation}
\end{lem}
%%%%%%%%%%%%%%%%%%%%%%%%%%%%%%%%%%%%%%%%%%%%%%%%%%%%%%%%%%%%%%%%

%%%%%%%%%%%%%%%%%%%%%%%%%%%%%%%%%%%%%%%%%%%%%%%%%%%%%%%%%%%%%%%%
\begin{proof}
The proof is due to an induction with respect to $k$. 
In this proof, the symbol $\norm{\ldots}$ means the norm of $\sF$.

For $k=1$, it can be seen that 
$\norm{(\Nf-\pi(K)-\xi_K)\Phi} \leq \norm{\Nf\Phi}+\norm{\pi(K)\Phi}+|\xi_K| \norm{\Phi}$.
Since $\norm{\pi(K)\Phi} \leq C |\xi_K|^{\frac{1}{2}} \norm{(\Nf+\one)^{\frac{1}{2}} \Phi}$,
\eqref{polyb} follows with $P_1(\tau) = 1+(C^2+\tau) + \tau$.

Next we suppose that \eqref{polyb} is true for $k=1,\ldots,n$. Then we have
\begin{align*}
  \norm{(\Nf-\pi(K)-\xi_K)^{n+1}\Phi}
  \leq &
   \norm{(\Nf-\pi(K)-\xi_K)^n \Nf\Phi}
   + \norm{(\Nf-\pi(K)-\xi_K)^n \pi(K) \Phi} \\
  & + \norm{(\Nf-\pi(K)-\xi_K)^n \xi_K \Phi}.
\end{align*}
By the induction hypothesis, it can be seen that
\begin{align*}
  \norm{(\Nf-\pi(K)-\xi_K)^n \Nf \Phi}      & \leq P_n(|\xi_K|) \norm{(\Nf+\one)^{n+1}\Phi}, \\
  \norm{(\Nf-\pi(K)-\xi_K)^n \xi_K \Phi}  & \leq P_n(|\xi_K|) |\xi_K| \norm{(\Nf+\one)^n \Phi}, \\
  \norm{(\Nf-\pi(K)-\xi_K)^n \pi(K) \Phi} & \leq P_n(|\xi_K|) \norm{(\Nf+\one)^n \pi(K) \Phi}.
\end{align*}
By a simple computation, we have 
\begin{align*}
  (\Nf+1)\pi(K)(\Nf+\one)^{-1} 
  & = \pi(K)+[\Nf,\pi(K)](\Nf+\one)^{-1} \\
  & = \pi(K) + iA(K) (\Nf+\one)^{-1},
\end{align*}
and hence the operator norm of $(\Nf+\one)^n \pi(K) (\Nf+\one)^{-(n+1)}$ can be estimated as 
\begin{align*}
 & \norm{(\Nf+\one)^n \pi(K) (\Nf+\one)^{-(n+1)}} \\
 & \leq 
   \norm{(\Nf+\one)^{n-1} \pi(K) (\Nf+\one)^{-n}}
   + \norm{ (\Nf+\one)^{n-1} A(K) (\Nf+\one)^{-(n+1)}} \\
 & \leq 
   \norm{ (\Nf+\one)^{n-1} \pi(K) (\Nf+\one)^{-n}} 
   + \norm{ (\Nf+\one)^{n-1} A(K) (\Nf+\one)^{-n}} \\
 & \leq \ldots 
  \leq 2^{n-1} C \norm{\pi(K)(\Nf+\one)^{-1}}
        +2^{n-1} C \norm{A(K)(\Nf+\one)^{-1}} 
         \leq 2^n C|\xi_K|^{\frac{1}{2}}.
\end{align*}
Thus we have
\begin{align*}
  \norm{(\Nf-\pi(K)-\xi_K)^{n+1}\Phi} 
  \leq
  P_n(|\xi_K|) ( 1 + |\xi_K| + 2^n(C^2+|\xi_K|) ) \norm{(\Nf+\one)^{n+1}\Phi}
\end{align*}
and the inequality \eqref{polyb} follows with 
$P_{n+1}(\tau) = P_n(\tau) (1+\tau + 2^n(C^2+\tau))$.
\end{proof}
%%%%%%%%%%%%%%%%%%%%%%%%%%%%%%%%%%%%%%%%%%%%%%%%%%%%%%%%%%%%%%%%

%%%%%%%%%%%%%%%%%%%%%%%%%%%%%%%%%%%%%%%%%%%%%%%%%%%%%%%%%%%%%%%%
\begin{lem}{\label{bd nen}}
 Assume (A1) and (A2). Let $n\in\NN$ be arbitrary.
Then, for any $\Psi \in \dom(\Nf^n)$ and $t\geq 0$, we have
$e^{-tT_\bA}\Psi \in \dom(\Nf^n)$ and 
\begin{align*}
 \norm{ \Nf^n e^{-tT_\bA}(\Nf+\one)^{-n} } \leq C_n (t^n+1)
\end{align*}
for some constant $C_n>0$.
\end{lem}
%%%%%%%%%%%%%%%%%%%%%%%%%%%%%%%%%%%%%%%%%%%%%%%%%%%%%%%%%%%%%%%%
\begin{proof}
It is enough to show that
\begin{align}
  \big| \inner{\Nf^n\Phi}{e^{-\frac{t}{2} T_\bA} \Psi} \big| \leq C \norm{\Phi},
  \qquad    \Phi \in \Hfin,  \label{invN}
\end{align}
with $C = C_n(t^n+1)\norm{(\Nf+\one)^n \Psi}$.
By the Feynman-Kac formula (Proposition \ref{fk}), the equivalence $\Pi(K) \cong \pi(K)$ and \eqref{throughN}, we have
\begin{align*}
  \big| \inner{\Nf^n\Phi}{e^{-\frac{t}{2}T_\bA} \Psi} \big| 
  & = \Big| \int_{\RR^3} \EE^\bx \big[ \inner{\cN^n \Phi(B_0)}{ e^{-i\cA(K)} \Psi(B_t)}_{L^2(Q)} \big] d\bx \Big| \\
  & = \Big| \int_{\RR^3} \EE^\bx\big[ 
      \inner{ \Phi(B_0) }{ e^{-i\cA(K)}( \cN-\Pi(K)-\xi_K)^n\Psi(B_t)}_{L^2(Q)} \big] d\bx \Big|.
\end{align*}
By Lemma \ref{thB}, we have
\begin{align}
 \big| \inner{\Nf^n\Phi }{e^{-\frac{t}{2}T_\bA} \Psi} \big| 
 \leq 
  \int_{\RR^3} \norm{\Phi(\bx)}_{L^2(Q)} \,
  \EE^\bx[P_n(|\xi_K|)^2]^{\frac{1}{2}} ~ 
  \EE^\bx[ \norm{ (\cN+\one)^n \Psi(B_t) }_{L^2(Q)}^2]^{\frac{1}{2}} d\bx. \label{esti3}
\end{align}
By the Burkholder-Davis-Gundy inequality \cite[Theorem 4.6]{hir00}
\begin{align*}
  \EE^\bx[ |\xi_K|^m] \leq c_m t^m \norm{ \phiome}^m, \qquad m\in\NN
\end{align*}
holds with some constant $c_m$ independent of $\bx$.
Then we get $\EE^\bx [ P_n(|\xi_K|)^2 ]^{\frac{1}{2}} < C_n(t^n+1)$ for some $C_n>0$, and 
hence the right-hand side of \eqref{esti3} is bounded by
\begin{align*}
& C_n(t^n+1) \int_{\RR^3} \norm{\Phi(\bx)}_{L^2(Q)} \EE^\bx[\norm{(\cN+\one)^n \Psi(B_t)}_{L^2(Q)}^2]^{\frac{1}{2}} d\bx \\
& \leq   C_n(t^n+1) \norm{\Phi} \int_{\RR^3} \EE^\bx[\norm{(\cN+\one)^n \Psi(B_t)}_{L^2(Q)}^2]^{\frac{1}{2}} d\bx 
 = C_n(t^n+1) \norm{\Phi} \norm{(\Nf+\one)^n \Psi}.
\end{align*}
Hence the proof is complete. 
\end{proof}
%%%%%%%%%%%%%%%%%%%%%%%%%%%%%%%%%%%%%%%%%%%%%%%%%%%%%%%%%%%%%%%%
Set %Let us set 
\begin{align*}
 R_s = (T_\bA + s)^{-1}.
\end{align*}
%%%%%%%%%%%%%%%%%%%%%%%%%%%%%%%%%%%%%%%%%%%%%%%%%%%%%%%%%%%%%%%%
\begin{lem}{\label{bd nrn}}
Assume (A1) and (A2). Let $n\in \NN$ and $s>0$. Then
it follows that 
$\ran(R_s (\Nf^n+\one)^{-1}) \subset \dom(\Nf^n)$, and 
\begin{align}
 \norm{ \Nf^n R_s (\Nf^n+\one)^{-1} }
 \leq  C_n( s^{-n-1} + s^{-1}) \label{bd nrn1}
\end{align}
holds for some $C_n>0$.
\end{lem}
%%%%%%%%%%%%%%%%%%%%%%%%%%%%%%%%%%%%%%%%%%%%%%%%%%%%%%%%%%%%%%%%
%
%%%%%%%%%%%%%%%%%%%%%%%%%%%%%%%%%%%%%%%%%%%%%%%%%%%%%%%%%%%%%%%%
\begin{proof}
Using the formula $(A+s)^{-1} = \int_0^\infty e^{-t(A+s)}dt$,
we have, for any $\Phi\in\Hfin$ and $\Psi\in \dom(\Nf)$,
\begin{align*}
 | \inner{\Nf^n \Phi}{R_s \Psi}|
 \leq 
 \int_0^\infty e^{-ts} \norm{\Phi}
 \norm{ \Nf^n e^{-tT_\bA} (\Nf^n +\one)^{-1} }
 \norm{(\Nf^n + \one)\Psi} dt.
\end{align*}
By  Lemma \ref{bd nen}, we have
\begin{align*}
 | \inner{\Nf^n \Phi}{R_s \Psi}|
 \leq \int_0^\infty e^{-ts} C_n(t^n + 1) \norm{\Phi} \norm{(\Nf^n+\one)\Psi}dt.
\end{align*}
Thus \eqref{bd nrn1} follows.
\end{proof}
%%%%%%%%%%%%%%%%%%%%%%%%%%%%%%%%%%%%%%%%%%%%%%%%%%%%%%%%%%%%%%%%%

We set $$T_{\bA,M}= T_{\bA}+M^2.$$
Note that $\dom(\sqrt{T_{\bA,M}})=\dom( \sqrt{T_\bA} )$, 
since $\sqrt{T_{\bA,M}}-\sqrt{T_{\bA}}$ is bounded. 
%%%%%%%%%%%%%%%%%%%%%%%%%%%%%%%%%%%%%%%%%%%%%%%%%%%%%%%%%%%%%%%%
\begin{lem}{\label{nsqn}}
Assume (A1) and (A2). Let $M>0$. Then
$T_{\bA,M}^{-\frac{1}{2}}\Psi \in \dom(\Nf)$ for any $\Psi \in \dom(\Nf)$, and 
\begin{align}
 \norm{\Nf T_{\bA,M}^{-\frac{1}{2}}(\Nf+1)^{-1}} 
 \leq 
 C_1\frac{1+2M^2}{2M^3}, \label{yy245}
\end{align}
where $C_1$ is the  constant in Lemma \ref{bd nrn}.
\end{lem}
%%%%%%%%%%%%%%%%%%%%%%%%%%%%%%%%%%%%%%%%%%%%%%%%%%%%%%%%%%%%%%%%
\begin{proof}
  By the integral expression of $T_{\bA,M}^{-\frac{1}{2}}$,
\begin{align*}
  T_{\bA,M}^{-\frac{1}{2}} = \frac{2}{\pi} \int_0^\infty  R_{\lambda^2+M^2} d\lambda,
\end{align*}
we have
\begin{align*}
  |\biginner{\Nf\Phi}{T_{\bA,M}^{-\frac{1}{2}}\Psi}|
  \leq \frac{2}{\pi} \int_0^\infty \norm{\Phi} \norm{\Nf R_{\lambda^2+M^2} \Psi}d\lambda.
\end{align*}
By Lemma \ref{bd nrn}, we have
\begin{align*} 
 \leq \frac{2C_1}{\pi} \norm{\Phi} \norm{(\Nf+1)\Psi} \int_0^\infty 
 ((\lambda^2+M^2)^{-2}+(\lambda^2+M^2)^{-1}) d\lambda. 
\end{align*}
Therefore $T_{\bA,M}^{-\frac{1}{2}}\Psi \in \dom(\Nf)$ and \eqref{yy245} hold.
\end{proof}

%%%%%%%%%%%%%%%%%%%%%%%%%%%%%%%%%%%%%%%%%%%%%%%%%%%%%%%%%%%%%%%%
\begin{lem}{\label{sq ta dn}}
Assume (A1) and (A2). Then (1) and (2) follow: 
\begin{itemize}
\item[\rm (1)] 
  For all $\Psi \in \dom(\Nf T_\bA) \cap \dom(\Nf) \cap \dom(\Nf T_\bA^2)$, 
  $T_\bA^{\frac{3}{2}}\Psi \in \dom(\Nf)$ and the bound 
\begin{align*}
  \norm{\Nf T_\bA^{\frac{3}{2}} \Psi}
  \leq C(\norm{\Nf T_\bA\Psi} + \norm{(\Nf+1)\Psi} + \norm{(\Nf+1)T_\bA^2\Psi})
\end{align*}
holds for some $C$ independent of $\Psi$.
\item[\rm (2)] For any $\Psi \in \Hfin$,
\begin{align*}
  \limsup_{M\to +0}\norm{\Nf T_\bA^2 T_{\bA,M}^{-\frac{1}{2}}\Psi} < \infty.
\end{align*}
\end{itemize}
\end{lem}
%%%%%%%%%%%%%%%%%%%%%%%%%%%%%%%%%%%%%%%%%%%%%%%%%%%%%%%%%%%%%%%%
\begin{proof}
By the integral expression of $T_\bA^{\frac{1}{2}}$, we have, for any $\Phi\in \Hfin$,
\begin{align*}
  \big| \biginner{ \Nf\Phi}{T_\bA^{\frac{3}{2}}\Psi} \big|
  \leq \frac{2}{\pi} \int_0^1 \big|\! \inner{\Nf\Phi}{ R_{\lambda^2} T_\bA^2 \Psi}\!\big| d\lambda 
  + \frac{2}{\pi}\int_1^\infty \big|\!\inner{\Nf\Phi}{ R_{\lambda^2} T_\bA^2 \Psi}\!\big| d\lambda.
\end{align*}
First we estimate the integral $\int_0^1 \ldots d\lambda$.
Since $T_\bA R_{\lambda^2} = \one - \lambda^2 R_{\lambda^2}$, we have
\begin{align*}
 \inner{\Nf\Phi}{ R_{\lambda^2} T_\bA^2 \Psi}
 & = \inner{\Nf\Phi}{T_\bA\Psi} - \lambda^2 \inner{\Nf\Phi}{R_{\lambda^2} T_\bA\Psi} \\
 & = \inner{\Nf\Phi}{T_\bA\Psi} - \lambda^2 \inner{\Nf\Phi}{(\one -\lambda^2 R_{\lambda^2}) \Psi},
\end{align*}
and hence
\begin{align*}
 \int_0^1 \left|\inner{\Nf\Phi}{ R_{\lambda^2}T_\bA^2 \Psi}\right| d\lambda 
\leq
  \norm{\Phi} \norm{\Nf T_\bA \Psi}
  + \int_0^1 \!\!\! \lambda^2 \norm{\Phi} \norm{\Nf\Psi} d\lambda 
  + \int_0^1 \!\!\! \lambda^4 \norm{\Phi} \norm{\Nf R_{\lambda^2}\Psi} d\lambda. 
\end{align*}
By  Lemma \ref{bd nrn}, we see that the last integral becomes finite and the bound
\begin{align*}
  \int_0^1 |\inner{\Nf\Phi}{R_{\lambda^2}T_\bA\Psi}| d\lambda 
 \leq C \norm{\Phi} \big( \norm{\Nf T_\bA \Psi} +  \norm{(\Nf +\one)\Psi} \big)
\end{align*}
holds for some $C>0$.
Next we consider the second part $\int_1^\infty d\lambda $.
By Lemma \ref{bd nrn} again, we get the bound
\begin{align*}
 \frac{2}{\pi} \int_1^\infty \big|\inner{\Nf \Phi}{R_{\lambda^2}T_\bA^2 \Psi} \big| d\lambda 
 & \leq 
 \frac{2}{\pi} \norm{\Phi} \int_1^\infty 
 C_1 (\lambda^{-4}+\lambda^{-2}) \norm{(\Nf +\one)T_\bA^2 \Psi} d\lambda   \\
 & = C \norm{\Phi} \norm{(\Nf +\one) T_\bA^2 \Psi}
\end{align*}
for some $C>0$. Since $\Phi\in\Hfin$ is arbitrary, these inequalities imply that
$T_\bA^{\frac{3}{2}}\Psi\in\dom(\Nf )$ and
\begin{align*}
  \norm{\Nf T_\bA^{\frac{3}{2}}\Psi}
  \leq C(\norm{\Nf T_\bA\Psi} + \norm{(\Nf +1)\Psi} + \norm{(\Nf +1)T_\bA^2\Psi})
\end{align*}
for some $C>0$. This shows (1).
The proof of (2) is completely similar to the proof of (1).
By Lemma \ref{dom1}, $\Hfin \subset \dom(\Nf T_\bA)\cap\dom(\Nf)\cap \dom(\Nf T_\bA^2)$.
Thus  as above, one can similarly show that 
\begin{align*}
  \sup_{0<M<1} \norm{\Nf  T_\bA^2 T_{\bA,M}^{-\frac{1}{2}}\Psi}
  \leq C(\norm{\Nf T_\bA\Psi} + \norm{(\Nf +1)\Psi} + \norm{(\Nf +1)T_\bA^2\Psi}),
\end{align*}
where $C$ is a constant independent of $\Psi$ and $M$. Thus (2) holds.
\end{proof}

We are in the position to  prove Proposition \ref{dom0}.
%%%%%%%%%%%%% proof of the dom0
\begin{proof}[Proof of Proposition \ref{dom0}:]
  Let $\Psi\in\Hfin$. Set $T=T_\bA$ and $T_M=T_{\bA,M}$ for simplicity.
We will show that 
\begin{align}
  \limsup_{M\to 0}\norm{\Nf^{\frac{1}{2}}T_M^{-\frac{1}{2}}T\Psi} < \infty.   \label{m0lim}
\end{align}
By Lemma \ref{dom1}, we have $\Psi \in \dom(T^2)$, in particular $\Psi \in \dom(T^{\frac{3}{2}})$.
Since $TT_M^{-\frac{1}{2}}\Psi \in \dom(T)$, there exists a sequence
$\{\Phi_j\}_j \subset \Hfin$, such that
$\Phi_j \to TT_M^{-\frac{1}{2}}\Psi$ and $T\Phi_j \to T^2T_M^{-\frac{1}{2}}\Psi$ as
$j\to \infty$. Then we have
\begin{align}
 \norm{\Nf^{\frac{1}{2}}T_M^{-\frac{1}{2}}T\Psi}^2
& = \biginner{TT_M^{-\frac{1}{2}}\Psi}{\Nf TT_M^{-\frac{1}{2}}\Psi}     
 = \lim_{j\to\infty} \biginner{\Phi_j}{\Nf TT_M^{-\frac{1}{2}}\Psi} \notag \\
& = \lim_{j\to\infty} \biginner{([T,\Nf]+\Nf T)\Phi_j}{T_M^{-\frac{1}{2}}\Psi}. \label{366}
\end{align}
The commutator $[\Nf ,T]$ can be computed as follows
\begin{align*}
 [\Nf ,T] 
 %  & = (\bp-\bA(\bx))\cdot[\Nf ,\bp-\bA(\bx)] + [\Nf ,\bp-\bA(\bx)]\cdot(\bp-\bA(\bx)) \\
 & = i(\bp-\bA(\bx))\cdot \boldsymbol{\pi} +i\boldsymbol{\pi} \cdot(\bp-\bA(\bx)),
\end{align*}
where $\boldsymbol{\pi}=(\pi_1,\pi_2,\pi_3)$ is defined by
\begin{align*}
  \pi_\mu = i[\Nf ,A_\mu(\bx)]
          = \frac{i}{\sqrt{2}} \big(-a(\overline{g_\mu(\bx)})+a^\dag(g_\mu(\bx)) \big),
\end{align*}
with $g_\mu(\bx) = e_\mu \phiome e^{-i\bk\cdot\bx} \in W$.
Since $\sum_{\mu=1}^3 [A_\mu(\bx), \pi_\mu] = 2i\norm{\phiome}^2$, we have
\begin{align*}
  [\Nf ,T] & = 2i\boldsymbol{\pi} \cdot(\bp-\bA(\bx)) +2\norm{\phiome}^2.
\end{align*}
Thus  \eqref{366} becomes
\begin{align*}
&  \lim_{j\to\infty} \Big(-2i \biginner{ \Phi_j}{\boldsymbol{\pi}\cdot(\bp-\bA(\bx))T_M^{-\frac{1}{2}}\Psi}
  - 2\norm{\phiome}^2\biginner{\Phi_j}{T_M^{-\frac{1}{2}}\Psi}
 + \biginner{T\Phi_j}{\Nf T_M^{-\frac{1}{2}}\Psi} \Big) \notag \\
& = \!-2i \biginner{TT_M^{-\frac{1}{2}}\Psi}{\boldsymbol{\pi}\cdot(\bp-\bA(\bx))T_M^{-\frac{1}{2}}\Psi}
 \!-\! 2\norm{\phiome}^2\biginner{TT_M^{-\frac{1}{2}}\Psi}{T_M^{-\frac{1}{2}}\Psi}
  \!+\! \biginner{T^2T_M^{-\frac{1}{2}}\Psi}{\Nf T_M^{-\frac{1}{2}}\Psi} \notag \\
& \leq -2i \biginner{TT_M^{-\frac{1}{2}}\Psi}{\boldsymbol{\pi}\cdot(\bp-\bA(\bx))T_M^{-\frac{1}{2}}\Psi}
  + \biginner{T^2T_M^{-\frac{1}{2}}\Psi}{\Nf T_M^{-\frac{1}{2}}\Psi}.
\end{align*}
Hence, by  the Schwarz inequality, we have
\begin{align*}
&\norm{\Nf^{\frac{1}{2}}T_M^{-\frac{1}{2}}T\Psi}^2 \\
&\leq 2\Big( \sum_{\mu=1}^3 \norm{\pi_\mu TT_M^{-\frac{1}{2}}\Psi}^2 \Big)^{\frac{1}{2}}
  \Big(\sum_{\mu=1}^3\norm{(p_\mu-A_\mu(\bx))T_M^{-\frac{1}{2}}\Psi}^2\Big)^{\frac{1}{2}} 
  + \norm{\Nf T^2T_M^{-\frac{1}{2}}\Psi} \norm{T_M^{-\frac{1}{2}}\Psi}.
\end{align*}
Noting $\sum_{\mu=1}^3(p_\mu-A_\mu(\bx))^2=T$ and 
\begin{align*}
\sum_{\mu=1}^3 \norm{\pi_\mu \Phi}^2 \leq 4\norm{\phiome}^2 \norm{(\Nf+1)^{\frac{1}{2}}\Phi}^2
\end{align*}
for $\Phi\in \dom(\Nf^{\frac{1}{2}})$, we have the bound
\begin{align}
 \norm{\Nf^{\frac{1}{2}}T_M^{-\frac{1}{2}}T\Psi}^2 
 \leq 4 \norm{\phiome} \norm{(\Nf+1)^{\frac{1}{2}} TT_M^{-\frac{1}{2}}\Psi} \norm{\Psi}
  + \norm{\Nf T^2T_M^{-\frac{1}{2}}\Psi} \norm{T_M^{-\frac{1}{2}}\Psi}. \label{375}
\end{align}
By Lemma \ref{sq ta dn}, we have 
\begin{align}
  \limsup_{M\to 0}\norm{\Nf T^2T_M^{-\frac{1}{2}}\Psi} < \infty,\qquad
  \limsup_{M\to 0}\norm{(\Nf+1)^{\frac{1}{2}} T^2T_M^{-\frac{1}{2}}\Psi} < \infty. \label{376}
\end{align}
On the other hand, since $\Psi\in \dom(|\bx|)$, by Lemma \ref{diamag}, 
we have $\Psi \in \dom(T^{-\frac{1}{2}})$ and 
\begin{align}
   \lim_{M\to 0} \norm{T_M^{-\frac{1}{2}}\Psi} = \norm{T^{-\frac{1}{2}}\Psi} 
   \leq 2\norm{|\bx|\Psi} < \infty.  \label{377}
\end{align}
Therefore, from \eqref{375}--\eqref{377}, we conclude that
\eqref{m0lim} holds.
By Lemma \ref{dom1} $T\Psi \in \dom(\Nf)$, and hence
$TT_M^{-\frac{1}{2}}\Psi = T_M^{-\frac{1}{2}}T\Psi \in \dom(\Nf)$ by Lemma \ref{nsqn}.
Thus $\Nf^{\frac{1}{2}}T_M^{-\frac{1}{2}}T\Psi\in \sH$.
By  \eqref{m0lim}, for any $\Phi\in\Hfin$, we see  that 
\begin{align*}
  \big|\biginner{T^{\frac{1}{2}}\Psi}{\Nf^{\frac{1}{2}}\Phi}\big|
 &= \lim_{M\to 0} \big|\biginner{TT_M^{-\frac{1}{2}}\Psi}{\Nf^{\frac{1}{2}}\Phi}\big| = \lim_{M\to 0} \big|\biginner{\Nf^{\frac{1}{2}}TT_M^{-\frac{1}{2}}\Psi}{\Phi}\big| \\
 &\leq \Big(\limsup_{M\to 0} \norm{\Nf^{\frac{1}{2}}TT_M^{-\frac{1}{2}}\Psi} \Big) \norm{\Phi}.
\end{align*}
Since $\Hfin$ is a core for $\Nf^{\frac{1}{2}}$, the above bound implies
$T^{\frac{1}{2}}\Psi \in \dom((\Nf^{\frac{1}{2}})^*)=\dom(\Nf^{\frac{1}{2}})$, which completes the proof
of Lemma \ref{dom0}.
\end{proof}
%%%%%%%%%%%%%%%%%%%%%%%%%%%%%%%%%%%%%%%%%%%%%%%%%%%%%%%%%%%%%%%%%

\section{Singular and non-local pull-through formulae} \label{sec:ptf}
Throughout we assume that (A1)--(A4) hold.
For $0<m<m_0$, recall that $\Phi_m$ is the normalized ground state of $H_m$.
%We define the annihilation operator kernel $a(k)$ as follows.
For each function $\Psi^{(n+1)} \in \stensor^{n+1} W$, 
the map $\RR^3\times \{1,2\} \ni k \mapsto \Psi^{(n+1)}(k,\ldots)$ is a 
$\stensor^n W$-valued function, and 
\begin{align*}
 \int \norm{\Psi^{(n+1)}(k,\ldots)}_{\stensor^n W}^2 dk 
 = \norm{\Psi^{(n+1)}}_{\stensor^{n+1} W }^2
\end{align*}
holds.
Thus  for each $\Psi \in \sF$ and almost every $k$, one can define the function
\begin{align*}
  (a\Psi)(k) = \left( \sqrt{n+1} \Psi^{(n+1)}(k,\cdot)\right)_{n=0}^\infty 
  \in \bigtimes_{n=0}^\infty \left( \stensor^n W \right),
\end{align*}
where $\bigtimes$ denotes the Cartesian product. %(see \cite{ge00}).
We write    $a(k) \Psi$ for 
$(a\Psi)(k)$.  
We  can check that $\Psi \in \dom(\Nf ^{\frac{1}{2}})$ if and only if
\begin{align*}
(1)\ & a(k)\Psi \in \sF \mbox{ a.e. } k,\\
(2)\ & \int \norm{a(k)\Psi}_\sF^2 dk < \infty.
\end{align*}
If $\Psi\in D(\Nf^{\frac{1}{2}})$, 
then 
\begin{align*}
&\norm{\Nf^{\frac{1}{2}}\Psi}_\sF^2 = \int \norm{a(k)\Psi}_\sF^2dk,\\
&
\inner{\Phi}{a(f)\Psi}_\sF = \int f(k)\inner{\Phi}{a(k)\Psi}_\sF dk 
\end{align*}
hold for all $\Phi \in \sF$ and $f\in W$.
For $\Psi\in\cH=L^2(\RR_\bx^3)\tensor \sF$, we can define $a(k)\Psi$ by 
$a(k)\Psi=\Psi(\bx,k,\ldots)$. 
In this section, we will establish the pull-through formula
\begin{align}
 a(k)\Phi_m =  \phiome(\bk)(H_m-E_m+\ome_m(\bk))^{-1}J(k) \Phi_m, 
 \label{pull-through}
\end{align}
where %$k=(\bk,j) \in \RR^3\times\{1,2\}$ and 
$J(k)$ is an operator valued function.
In the case of $M=0$, it is crucial to consider the operator domain
in the derivation of \eqref{pull-through}.

Let $f\in C_0^\infty(\RR^3\times\{1,2\})$ and $\Psi \in \Hfin$.
By Proposition \ref{dom0}, we have $T_\bA^{\frac{1}{2}} \Psi \in \dom(\Nf ^{\frac{1}{2}}) \subset \dom(a^\dag(f))$
and $a^\dag(f)\Psi \in \Hfin \subset \dom(H_m)$ follows.
From these facts, we can verify the following calculations.
\begin{align*}
& \inner {(H_m-E_m)\Psi}{a(\bar{f})\Phi_m}  = \inner{ a^\dag(f) (H_m - E_m)\Psi }{ \Phi_m } \\
& = \inner{ [ a^\dag(f), H_m - E_m] \Psi }{ \Phi_m }
    + \inner{ ( H_m - E_m ) a^\dag(f) \Psi }{ \Phi_m } 
    = \inner{ [ a^\dag(f), H_m ] \Psi }{ \Phi_m }.
\end{align*}
Since
\[
 [a^\dag(f), H_m ] = [a^\dag(f), \sqrt{T_\bA} ] -a^\dag(\ome_m f)
\]
holds on $\Hfin$, we have
\begin{align}
& \inner {(H_m-E_m)\Psi}{a(\bar{f})\Phi_m}            
 = 
 \biginner{ [a^\dag(f), \sqrt{T_\bA} ] \Psi}{ \Phi_m}
 - \inner{ \Psi }{ a(\ome_m \bar{f}) \Phi_m }         \notag \\
& = 
 \biginner{ \sqrt{T_\bA }  \Psi}{ a(\bar{f})\Phi_m} 
 - \biginner{ a^\dag(f)\Psi }{ \sqrt{T_\bA}\Phi_m } 
 - \inner{ \Psi }{ a(\ome_m \bar{f}) \Phi_m }       \notag \\
& = 
 \frac{2}{\pi} \int_0^\infty 
 \big( \inner{ T_\bA R_{t^2} \Psi }{ a(\bar{f})\Phi_m } 
   - \inner{ a^\dag(f)\Psi }{ T_\bA R_{t^2} \Phi_m } \big) dt 
   - \inner{ \Psi }{ a(\ome_m \bar{f}) \Phi_m }  \notag \\
& = 
 \frac{2}{\pi} \int_0^\infty
 \inner{ [a^\dag(f) , T_\bA R_{t^2}] \Psi}{ \Phi_m} dt 
 - \inner{ \Psi }{ a(\ome_m \bar{f}) \Phi_m }, \label{xx215}
\end{align}
where we used the formula:
\begin{align}
  \sqrt{S} = \frac{2}{\pi} \int_0^\infty \frac{S}{S+t^2} dt, \qquad S\geq 0. 
  \label{sqrt int}
\end{align}
%which is used with a spectral theorem.
We shall compute the commutator in the integrand of \eqref{xx215}. 
It  is enough to consider the case $t>0$.
Note that $R_{t^2} \Psi \in \dom(\Nf )$ by Lemma \ref{bd nrn}.
By $S/(S+t^2)=\one-t^2/(S+t^2)$ and the resolvent equation, we have
\begin{align*}
   \inner{ [a^\dag(f) , T_\bA R_{t^2}] \Psi}{ \Phi_m} 
= -t^2 \inner{ [a^\dag(f) , R_{t^2}] \Psi}{ \Phi_m} 
 = -t^2 \inner{ [T_\bA, a^\dag(f)] R_{t^2} \Psi}{ R_{t^2} \Phi_m}. 
\end{align*}
The commutator above is estimated as 
\begin{align*}
 [T_\bA, a^\dag(f)] 
& = (\bp-\bA(\bx)) \cdot [\bp-\bA(\bx), a^\dag(f)]
 + [\bp-\bA(\bx), a^\dag(f)] \cdot (\bp-\bA(\bx)) \\
& = - \sqrt{2} (\bp-\bA(\bx))
   \cdot \inner{ e^{-i\bk\cdot\bx} \be \phiome }{f}_{W},
\end{align*}
where $(e^{-i\bk\cdot\bx}\be \phiome)(\bk,j)=e^{-i\bk\cdot\bx} \phiome(\bk) (e_1(k), e_2(k), e_3(k))$.
Thus
\begin{align}
& \inner{ [a^\dag(f) , T_\bA R_{t^2}] \Psi}{ \Phi_m} = \sqrt{2} t^2
  \inner{ \inner{ e^{-i\bk\cdot\bx} \be \phiome }{f} R_{t^2} \Psi }
        { (\bp-\bA(\bx)) R_{t^2} \Phi_m}   \notag \\
& = \sqrt{2} t^2 \int \overline{f(k)} \phiome(\bk)
  \inner{ e^{i\bk\cdot\bx} R_{t^2} \Psi }
        { \be(k) \cdot (\bp-\bA(\bx)) R_{t^2} \Phi_m} dk   \notag  \\
& = \sqrt{2} t^2 \int \overline{f(k)} \phiome(\bk)
  \inner{ e^{i\bk\cdot\bx} R_{t^2} \Psi }
        { V_{\be(k)} R_{t^2} \Phi_m } dk,  \label{xx221}
\end{align}
where, for $\bw\in\RR^3$, we introduced the operator
\begin{align}
   V_\bw = \bw \cdot (\bp-\bA(\bx)).   \label{def v}
\end{align}
Therefore, the first term in \eqref{xx215} becomes
\begin{align}
 \frac{2}{\pi} \int_0^\infty\!\!\!
  \inner{ [a^\dag(f) , T_\bA R_{t^2}] \Psi}{ \Phi_m} dt 
 = \frac{2\sqrt{2}}{\pi} \int_0^\infty \!\!\! t^2 dt 
    \int \overline{f(k)} \phiome(\bk)
    \inner{e^{i\bk\cdot\bx} R_{t^2} \Psi}{V_{\be(k)}R_{t^2}\Phi_m} 
    dk .  \label{xx222}
\end{align}
Although the iterated integral in \eqref{xx222}  converges, 
 the total integrability is not clear, especially around $t=0$.
In order to use Fubini's lemma, we have to show the total 
integrability of \eqref{xx222}.

We show several properties on $V_\bw$ in the next lemma.
%%%%%%%%%%%%%%%%%%%%%%%%%%%%%%%%%%%%%%%%%%%%%%%%%%%%%%%%%%%%%%%%
\begin{lem}{\label{prop of v}}
  Assume (A1) and (A2). Then, for any $\bw\in\RR^3$, $V_\bw$
is essentially self-adjoint on $\Hfin$.
We use the same symbol for its closure.
Moreover, the following hold:
\begin{enumerate}
\item[\rm (1)] If $\Psi\in \dom(T_\bA^{\frac{1}{2}})$, then $\Psi \in \dom(V_\bw)$ and 
$\norm{V_\bw \Psi}\leq |\bw| \norm{T_\bA^{\frac{1}{2}}\Psi}.$
\item[\rm (2)] If $\Psi\in \dom(T_\bA^{\frac{1}{4}})$, then $\Psi \in \dom(|V_\bw|^{\frac{1}{2}})$ and 
$\norm{|V_\bw|^{\frac{1}{2}} \Psi}\leq |\bw|^{\frac{1}{2}} \norm{T_\bA^{\frac{1}{4}}\Psi}.$
\item[\rm (3)]
 For all $\bk\in \RR^3$ with $(k_1,k_2)\neq (0,0)$, 
$V_{\be(k)}$ strongly commutes with $e^{-i\bk\cdot\bx}$.
\end{enumerate}
\end{lem}
%%%%%%%%%%%%%%%%%%%%%%%%%%%%%%%%%%%%%%%%%%%%%%%%%%%%%%%%%%%%%%%%
\begin{proof}
The essential self-adjointness follows from Nelson's commutator theorem with 
auxiliary operator $|\bp|^2+\Nf+\one$.
For $\Psi \in \dom(T_\bA^{\frac{1}{2}})$, by the Schwarz inequality,
\begin{align*}
 \norm{V_\bw \Psi}^2 
&\leq \sum_{\mu,\nu=1}^3 |w_\mu w_\nu|  |\inner{(p_\mu-A_\mu(\bx))\Psi}{(p_\nu-A_\nu(\bx))\Psi}| \\
&\leq \Big(\sum_{\mu=1}^3 |w_\mu|  \norm{(p_\mu-A_\mu(\bx))\Psi} \Big)^2 
\leq |\bw|^2  \sum_{\mu=1}^3\norm{(p_\mu-A_\mu(\bx))\Psi}^2,
\end{align*}
which implies (1).
The statement (2) can be derived from the L\"owner-Heinz inequality \cite[Theorem 2]{k52}.
Finally  we prove (3). Note that $e^{-i\bk\cdot\bx}$ is a unitary operator.
Noting $\bk\cdot\be(k)=0$, we can  show that
\begin{align*}
  e^{i\bk\cdot\bx} \be(k)\cdot(\bp-\bA(\bx)) e^{-i\bk\cdot\bx} = 
\be(k)\cdot(\bp-\bk-\bA(\bx))
= \be(k)\cdot(\bp-\bA(\bx))
\end{align*}
on $\Hfin$. 
Taking the closure on both sides, we have $e^{i\bk\cdot\bx}V_{\be(k)}e^{-i\bk\cdot\bx}=V_{\be(k)}$.
Thus (3) is proven.
\end{proof}
%%%%%%%%%%%%%%%%%%%%%%%%%%%%%%%%%%%%%%%%%%%%%%%%%%%%%%%%%%%%%%%%

The next lemma shows that the integral in \eqref{xx222} is absolutely convergent.
%%%%%%%%%%%%%%%%%%%%%%%%%%%%%%%%%%%%%%%%%%%%%%%%%%%%%%%%%%%%%%%%
\begin{lem}{\label{fubini00}}
For $k=(\bk,j)\in \RR^3\times\{1,2\}$ with $(k_1,k_2)\neq (0,0)$ and 
$\Psi, \Phi \in \sH$,
the bound
\begin{align}
 \int_0^\infty \big|\inner{e^{i\bk\cdot\bx} R_{t^2} \Psi}{V_{\be(k)} R_{t^2}\Phi}\big| t^2 dt 
 \leq \frac{\pi}{4} \norm{\Psi} \norm{\Phi}   \label{fubini01}
\end{align}
holds, and 
$
  t^2 \big| f(k)\phiome(\bk)\inner{e^{i\bk\cdot\bx} R_{t^2} \Psi}{V_{\be(k)}R_{t^2}\Phi}\big|
$
is integrable in $(k,t) \in (\RR^3\times\{1,2\}) \times [0,\infty)$.
\end{lem}
%%%%%%%%%%%%%%%%%%%%%%%%%%%%%%%%%%%%%%%%%%%%%%%%%%%%%%%%%%%%%%%%
\begin{proof}
Note that $R_{t^2}\Phi, R_{t^2}\Psi \in \dom(V_{\be(k)})$ for all $t>0$
and $\Psi,\Phi \in \sH$.
For $t>0$ and $\bk \in \RR^3\backslash L_{12}$, we have
\begin{align}
&  \big| \inner{e^{i\bk\cdot\bx}R_{t^2}\Psi}{V_{\be(k)}R_{t^2}\Phi}\big|
 = \big| \inner{|V_{\be(k)}|^{\frac{1}{2}} e^{i\bk\cdot\bx}R_{t^2}\Psi}
              {\sgn(V_{\be(k)})|V_{\be(k)}|^{\frac{1}{2}}R_{t^2}\Phi}\big| \notag \\
& \leq \norm{|V_{\be(k)}|^{\frac{1}{2}} e^{i\bk\cdot\bx}R_{t^2}\Psi} \norm{V_{\be(k)}|^{\frac{1}{2}}R_{t^2}\Phi} \notag 
 \leq \norm{|V_{\be(k)}|^{\frac{1}{2}} R_{t^2}\Psi}  \norm{V_{\be(k)}|^{\frac{1}{2}}R_{t^2}\Phi} \notag \\
& \leq \norm{T_\bA^{\frac{1}{4}} R_{t^2}\Psi} \norm{T_\bA^{\frac{1}{4}}R_{t^2}\Phi}, \label{fubini02}
\end{align}
where we used Lemma \ref{prop of v} and the fact that $\be(k)$ is a normalized vector.  
Thus 
\begin{align*}
&\int_0^\infty \big| \inner{e^{i\bk\cdot\bx}R_{t^2}\Psi}{V_{\be(k)}R_{t^2}\Phi_m}\big| t^2 dt \\
& \leq \Big( \int_0^\infty \norm{T_\bA^{\frac{1}{4}}R_{t^2}\Phi}^2 dt \Big)^{\frac{1}{2}}
\Big( \int_0^\infty \norm{T_\bA^{\frac{1}{4}}R_{t^2}\Psi}^2 t^2 dt \Big)^{\frac{1}{2}}.
\end{align*}
Since $\int_0^\infty \norm{T_\bA^{\frac{1}{4}}R_{t^2}\Psi}^2 t^2 dt  = (\pi/4)\norm{\Psi}^2$,
\eqref{fubini01} follows.
\end{proof}
As a consequence of Lemma \ref{fubini00}, we can apply Fubini's lemma to
\eqref{xx222}, and we have
\begin{align}
  \eqref{xx222}
 = \frac{2\sqrt{2}}{\pi} \int \overline{f(k)} \phiome(\bk)
dk \int_0^\infty \inner{e^{i\bk\cdot\bx} R_{t^2} \Psi}{V_{\be(k)}R_{t^2}\Phi_m} t^2dt. \label{xx436}
\end{align}
%%%%%%%%%%%%%%%%%%%%%%%%%%%%%%%%%%%%%%%%%%%%%%%%%%%%%%%%%%%%%%%%%
Thus we obtain the following result.
%%%%%%%%%%%%%%%%%%%%%%%%%%%%%%%%%%%%%%%%%%%%%%%%%%%%%%%%%%%%%%%%
\begin{cor}{\label{def of j}}
For each $k=(\bk,j)\in\RR^3\times\{1,2\}$ with $(k_1,k_2)\neq (0,0)$,
 the integral
\begin{align}
 J(k) 
 = \frac{2\sqrt{2}}{\pi} \int_0^\infty R_{t^2} e^{-i\bk\cdot\bx} 
    V_{\be(k)} R_{t^2} t^2 dt    \label{xx230}
\end{align}
defines a bounded operator on $\sH$ with the operator norm
\begin{align*}
  \norm{J(k)} \leq \frac{1}{\sqrt{2}}.
\end{align*}
\end{cor}
%%%%%%%%%%%%%%%%%%%%%%%%%%%%%%%%%%%%%%%%%%%%%%%%%%%%%%%%%%%%%%%%
\begin{proof}
This is a direct consequence of Lemma \ref{fubini00}.
\end{proof}
%%%%%%%%%%%%%%%%%%%%%%%%%%%%%%%%%%%%%%%%%%%%%%%%%%%%%%%%%%%%%%%%
Now we can state the main proposition in this section.
%%%%%%%%%%%%%%%%%%%%%%%%%%%%%%%%%%%%%%%%%%%%%%%%%%%%%%%%%%%%%%%%
\begin{prop}[Singular and non-local pull-through formula]
  Assume (A1)--(A4). For all $m>0$ and a.e.  $k=(\bk,j)\in\RR^3\times\{1,2\}$,
it follows that 
\begin{align}
  a(k)\Phi_m = \phiome(\bk)(H_m-E_m + \ome_m(\bk))^{-1} J(k) \Phi_m.  \label{puth}
\end{align}
\end{prop}
%%%%%%%%%%%%%%%%%%%%%%%%%%%%%%%%%%%%%%%%%%%%%%%%%%%%%%%%%%%%%%%%
\begin{proof}
Combining \eqref{xx436} and Corollary \ref{def of j}, we have the identity
\begin{align*}
& \int \overline{f(k)}\inner{(H_m-E_m)\Psi}{a(k)\Phi_m} dk
 + \int \overline{f(k)} \ome_m(\bk)\inner{\Psi}{a(k)\Phi_m} dk \\
& = \int \overline{f(k)} \phiome(\bk) \inner{\Psi}{ J(k) \Phi_m} dk
\end{align*}
for all $f\in C_0^\infty(\RR^3\times\{1,2\})$ and $\Psi\in \Hfin$.
Thus
\begin{align}
  \inner{(H_m-E_m+\ome_m(\bk))\Psi}{a(k)\Phi_m}
  = \phiome(\bk) \inner{ \Psi } { J(k) \Phi_m} \label{xx438}
\end{align}
holds for all $\Psi\in \Hfin$ and $k=(\bk,j)\in (\RR^3\times\{1,2\}) \backslash N_\Psi$
with some null sets $N_\Psi$.
Since $\Hfin$ is dense and we can take a countable dense subset $\cD$ of $\Hfin$,  
\eqref{xx438} holds for  $\Psi\in \cD$ for $k\in (\RR^3\times\{1,2\}) \backslash (\cup_{\Phi\in\cD}N_\Phi)$:
$$(H_m-E_m+\ome_m(\bk)){a(k)\Phi_m}= \phiome(\bk)  J(k) \Phi_m$$ for 
  $k\in (\RR^3\times\{1,2\}) \backslash (\cup_{\Phi\in\cD}N_\Phi)$. 
Therefore \eqref{puth} follows.
\end{proof}

\section{Photon number localization} \label{sec:pmdb}
Our goal in this section is to prove the following result.
%%%%%%%%%%%%%%%%%%%%%%%%%%%%%%%%%%%%%%%%%%%%%%%%%%%%%%%%%%%%%%%%
\begin{prop}
{\label{pdenb}}
 Assume (A1)--(A4). Let $0<m< m_0$. Then, there exists a constant $C>0$
 independent of $m$ such that 
\begin{align}
  \norm{a(k)\Phi_m}^2 \leq C \frac{|\phih(\bk)|^2}{\ome(\bk)}(1+|\bk|)^2
  \label{xy237}
\end{align}
for a.e. $k=(\bk,j)\in \RR^3\times\{1,2\}$.
\end{prop}
%%%%%%%%%%%%%%%%%%%%%%%%%%%%%%%%%%%%%%%%%%%%%%%%%%%%%%%%%%%%%%%%
We can show the uniform photon number localization of $\Phi_m$ as a corollary of Proposition \ref{pdenb}: 
\begin{cor}
\label{NN}
Assume (A1)--(A4). 
Then $\displaystyle \sup_{0<m<m_0}\|\Nf^{\frac{1}{2}}\Phi_m\|<\infty$.
\end{cor}
\begin{proof}
By  Corollary \ref{def of j}, we can have the bound
\begin{align}
 \norm{a(k)\Phi_m}^2
 \leq 
 |\phiome(\bk)|^2 \norm{(H_m-E_m+\ome_m(\bk))^{-1}}^2
 \norm{J(k)}^2 
 \leq \frac{|\hat\varphi(\bk)|^2}{2\ome(\bk)^{\frac{3}{2}}}. \label{IRdiv}
\end{align}
%On the other hand \eqref{IRdiv} decays however faster than \eqref{xy237} for large $|\bk|$. 
Combining \eqref{xy237} and \eqref{IRdiv}, we get the bound
\begin{align}\label{Nbound}
  \norm{a(k)\Phi_m}^2 \leq 
  \frac{|\phih(\bk)|^2}{2\ome(\bk)}\min\{2C(1+|\bk|^2), \ome(\bk)^{-\frac{1}{2}}\}.
\end{align}
By \eqref{Nbound} we have 
$$\|\Nf^{\frac{1}{2}}\Phi_m\|^2\leq \int_{\RR^3} 
\frac{|\phih(\bk)|^2}{2\ome(\bk)}\min\{2C(1+|\bk|^2), \ome(\bk)^{-\frac{1}{2}}\}d\bk<\infty$$
Take $\sup_{0<m<m_0}$ on both sides above. Thus the corollary follows.
\end{proof}
\begin{rem}
The right-hand side of \eqref{IRdiv}  has a singularity at $\bk=0$, and then 
the right-hand side of \eqref{IRdiv}
is not integrable if $\hat\varphi(0)\neq 0$.
This type of singularity 
 is often referred to as
 an infrared divergence.
\end{rem} 
To derive \eqref{xy237}
%In the proof of Proposition \ref{pdenb}, in order to avoid the infrared divergence, 
we use a method due to  \cite[p.214]{hir03} and \cite[(7.7)]{hhs05}. 
%instead of the Pauli-Fierz transformation.
We decompose $J(k)$ into three terms: 
\begin{align}
 J(k) = \frac{2\sqrt{2}}{\pi} \big( 
  L_1(k) \braket{\bx}^2 + L_2(k) \braket{\bx} + L_3(k) \big), \label{xy243}
\end{align}
where 
\begin{align*}
& L_1  = L_1(k) 
 = \int_0^1 R_{t^2} V_{\be(k)} (e^{-i\bk\cdot\bx}-1) R_{t^2} \braket{\bx}^{-2} t^2dt , \\
& L_2  = L_2(k) 
 = \int_1^\infty R_{t^2} V_{\be(k)} (e^{-i\bk\cdot\bx}-1) R_{t^2} \braket{\bx}^{-1}t^2 dt, \\
& L_3 = L_3(k) 
 = \int_0^\infty R_{t^2} V_{\be(k)} R_{t^2} t^2 dt.
\end{align*}
Note that the velocity operator $V_{\be(k)}$ commutes with $e^{-i\bk\cdot\bx}$.
%%%%%%%%%%%%%%%%%%%%%%%%%%%%%%%%%%%%%%%%%%%%%%%%%%%%%%%%%%%%%%%%
\subsection{Estimate on $L_1$}
In order to prove that $L_1(k)$ is bounded, we introduce an operator $Z$ by
\begin{align}
 Z = \int_0^1 \braket{\bx}^{-2} (t^2+|\bp|^2)^{-1} |\bx|^2 (t^2+|\bp|^2)^{-1} \braket{\bx}^{-2} t^3dt.   \label{def z}
\end{align}
%%%%%%%%%%%%%%%%%%%%%%%%%%%%%%%%%%%%%%%%%%%%%%%%%%%%%%%%%%%%%%%%
\begin{lem}{\label{bound z}}
 The operator $Z$ is non-negative, bounded and $\norm{Z}\leq 6$.
\end{lem}
%%%%%%%%%%%%%%%%%%%%%%%%%%%%%%%%%%%%%%%%%%%%%%%%%%%%%%%%%%%%%%%%
\begin{proof}
Since $Z$ is symmetric and non-negative, it is enough to show that
$|\inner{u}{Zu}|\leq C\norm{u}^2, u\in L^2(\RR^3)$ for some $C>0$.
We use the commutation relation: 
\begin{align*}
 x_\mu (t^2+|\bp|^2)^{-1} = (t^2+|\bp|^2)^{-1} x_\mu + \frac{-2ip_\mu}{(t^2+|\bp|^2)^2}.
\end{align*}
For $u\in L^2(\RR^3)$, we have
\begin{align*}
   |\inner{u}{Zu}| 
 = & \sum_{\mu=1}^3 \int_0^1 \norm{x_\mu (t^2+|\bp|^2)^{-1} \braket{\bx}^{-2} u}^2 t^3 dt \\
 = & \sum_{\mu=1}^3 \int_0^1 \norm{(t^2+|\bp|^2)^{-1} x_\mu \braket{\bx}^{-2} u}^2 t^3dt \\
   & + 4\Im \int_0^1 \inner{\bp\cdot \bx \braket{\bx}^{-2} u}{ (t^2+|\bp|^2)^{-3} \braket{\bx}^{-2} u} t^3dt \\
   & +\sum_{\mu=1}^3 \int_0^1 \norm{-2ip_\mu (t^2+|\bp|^2)^{-2} \braket{\bx}^{-2} u}^2 t^3 dt.
\end{align*}
Note that
\begin{align*}
& \int_0^1 \frac{t^3}{(t^2+|\bp|^2)^2} dt
 = \half\Big( \log(1+\frac{1}{|\bp|^2}) - \frac{1}{1+|\bp|^2} \Big)
 < \frac{1}{2|\bp|^2}, \\
& \int_0^1 \frac{t^3}{(t^2+|\bp|^2)^3} dt
 = \frac{1}{4|\bp|^2(1+|\bp|^2)^2}
 \leq \frac{1}{4|\bp|^2}, \\
& \int_0^1 \frac{t^3}{(t^2+|\bp|^2)^4} dt
 = \frac{1}{12|\bp|^4(1+|\bp|^2)^2} + \frac{1}{6|\bp|^2(1+|\bp|^2)^3}
\leq \frac{1}{12|\bp|^4}.
\end{align*}
Thus  we have
\begin{align*}
|\!\inner{u}{Zu}\!| 
&\!\leq\! \frac{1}{2} 
\sum_{\mu=1}^3 \norm{|\bp|^{-1} x_\mu \braket{\bx}^{-2} \!u}^2
      +\!\! \sum_{\mu=1}^3 \norm{x_\mu \braket{\bx}^{-2}\!u} \Bignorm{\frac{p_\mu}{|\bp|^2}\braket{\bx}^{-2}u} 
     \! + \!\frac{1}{3} \norm{|\bp|^{-1} \braket{\bx}^{-2} \!u}^2 \\
& \leq \frac{1}{2} \sum_{\mu=1}^3 \norm{|\bp|^{-1} x_\mu \braket{\bx}^{-2} \!u}^2
      + \norm{\braket{\bx}^{-1}u} \norm{ |\bp|^{-1} \!\braket{\bx}^{-2}u}
      + \frac{1}{3} \norm{|\bp|^{-1} \braket{\bx}^{-2} u}^2.
\end{align*}
By Hardy's inequality, we have
\begin{align*}
  |\inner{u}{Zu}| 
 & \leq 2 \norm{|\bx|^2 \braket{\bx}^{-2}u}^2 
   + 2\norm{\braket{\bx}^{-1} u} \norm{|\bx|\braket{\bx}^{-2}u}
   + \frac{4}{3} \norm{|\bx| \braket{\bx}^{-2}u}^2 \\
 & \leq \frac{16}{3}\norm{u}^2 \leq 6\norm{u}^2
\end{align*}
for all $u\in L^2(\RR^3)$. Then the proof is complete. 
\end{proof}
%%%%%%%%%%%%%%%%%%%%%%%%%%%%%%%%%%%%%%%%%%%%%%%%%%%%%%%%%%%%%%%%
\begin{lem}{ \label{bound l1} }
For every $k\in \RR^3\times\{1,2\}$, operator $L_1(k)$ is bounded and 
\begin{align}
  \norm{L_1(k)} \leq  2|\bk|.  \label{xx256}
\end{align}
\end{lem}
%%%%%%%%%%%%%%%%%%%%%%%%%%%%%%%%%%%%%%%%%%%%%%%%%%%%%%%%%%%%%%%%
\begin{proof}
 For any $\Psi,\Phi \in \sH$, we have
\begin{align*}
 |\inner{ \Psi}{ L_1(k) \Phi} | 
& \leq 
 \int_0^1 \norm{V_{\be(k)} R_{t^2}\Psi} 
 \norm{ \bk\cdot\bx R_{t^2}\braket{\bx}^{-2} \Phi} t^2 dt \\
& \leq 
 |\bk| \int_0^1 \norm{ T_\bA^{\frac{1}{2}} R_{t^2}\Psi} \norm{ |\bx| (|\bp|^2+t^2)^{-1} \braket{\bx}^{-2}|\Phi|} t^2 dt \\
& \leq 
 |\bk| \Big(\int_0^1 \norm{ T_\bA^{\frac{1}{2}} R_{t^2}\Psi}^2 tdt \Big)^{\frac{1}{2}}
 \Big( \int_0^1 \norm{ |\bx| (|\bp|^2+t^2)^{-1}\braket{\bx}^{-2}|\Phi|}^2 t^3dt \Big)^{\frac{1}{2}}.
\end{align*}
Here we used Lemma \ref{prop of v} and the diamagnetic inequality (Lemma \ref{diamag} (3)) for the second inequality,
 and the Schwarz inequality for the third inequality.
Since 
\begin{align*}
\left\|
  \int_0^1 \frac{T_\bA}{(T_\bA+t^2)^2} tdt \right\|= 
  \left\|
  \frac{1}{2(T_\bA+1)} \right\|\leq \frac{1}{2},
\end{align*}
we have
\begin{align*}
  |\inner{ \Psi}{ L_1(k) \Phi} | 
  \leq \frac{1}{\sqrt{2}}|\bk| \norm{\Psi} \inner{|\Phi|}{Z|\Phi|}^{\frac{1}{2}}.
\end{align*}
This estimate and Lemma \ref{bound z} imply \eqref{xx256}.
\end{proof}
%%%%%%%%%%%%%%%%%%%%%%%%%%%%%%%%%%%%%%%%%%%%%%%%%%%%%%%%%%%%%%%%
\subsection{Estimate on $L_2$}
We shall estimate $L_2(k)$.
Set 
$$T_{\bA-\bk}=(\bp+\bk-\bA(\bx))^2,\quad R_{t^2}(\bk)=(T_{\bA-\bk}+t^2)^{-1}.$$
We have identities:
\begin{align*}
  (e^{-i\bk\cdot\bx}-1) R_{t^2}
  & = R_{t^2}(\bk) (e^{-i\bk\cdot\bx}-1) + R_{t^2}(T_\bA-T_{\bA-\bk})R_{t^2}(\bk), \\
  T_\bA-T_{\bA-\bk}
  & = -2V_\bk - |\bk|^2.
\end{align*}
We then have 
\begin{align*}
  (e^{-i\bk\cdot\bx}-1) R_{t^2}
  = R_{t^2}(\bk) (e^{-i\bk\cdot\bx}-1)
    -2 R_{t^2} V_\bk R_{t^2}(\bk)
    - |\bk|^2 R_{t^2} R_{t^2}(\bk).
\end{align*}
According to above identity  
we decompose $L_2(k)$ into three terms:
\begin{align}
  L_2(k) = L_{21}(k) + L_{22}(k) + L_{23}(k), \label{xy266}
\end{align}
where
\begin{align*}
 L_{21}(k) & = \int_1^\infty R_{t^2} V_{\be(k)} R_{t^2}(\bk) (e^{-i\bk\cdot\bx}-1) \braket{\bx}^{-1} t^2dt,  \\
 L_{22}(k) &= -2 \int_1^\infty R_{t^2} V_{\be(k)} R_{t^2} V_\bk R_{t^2}(\bk) \braket{\bx}^{-1} t^2dt,  \\
 L_{23}(k) &= -|\bk|^2 \int_1^\infty R_{t^2} V_{\be(k)} R_{t^2} R_{t^2}(\bk) \braket{\bx}^{-1} t^2dt.
\end{align*}
In order to estimate $L_{21}$, we show the next lemma.
%%%%%%%%%%%%%%%%%%%%%%%%%%%%%%%%%%%%%%%%%%%%%%%%%%%%%%%%%%%%%%%%
\begin{lem}{ \label{bound tb}}
If $\Psi \in \dom(T_{\bA-\bk}^{\frac{1}{2}})$ and $\Phi\in \dom(T_{\bA-\bk}^{\frac{1}{4}})$, then
\begin{align}
&  \norm{V_{\be(k)}\Psi} \leq \norm{T_{\bA-\bk}^{\frac{1}{2}} \Psi},          \label{xx269} \\
&  \norm{|V_{\be(k)}|^{\frac{1}{2}}\Phi} \leq \norm{T_{\bA-\bk}^{\frac{1}{4}} \Phi}  \label{xx270}
\end{align}
hold for $k=(\bk,j)\in \RR^3\times\{1,2\}$.
\end{lem}
%%%%%%%%%%%%%%%%%%%%%%%%%%%%%%%%%%%%%%%%%%%%%%%%%%%%%%%%%%%%%%%%
\begin{proof}
Note that $\be(k)\perp \bk$ and $V_{\be(k)}=\be(k)\cdot(\bp+\bk-\bA(\bx))$ hold.
Thus  the proof is the same as that  of Lemma \ref{prop of v}.
\end{proof}
%%%%%%%%%%%%%%%%%%%%%%%%%%%%%%%%%%%%%%%%%%%%%%%%%%%%%%%%%%%%%%%%
\begin{lem}{\label{bound l21}}
For $k=(\bk,j)\in \RR^3\times\{1,2\}$, we have
\begin{align}
  \norm{L_{21}(k)}  \leq |\bk|. \label{xx271}
\end{align}
\end{lem}
%%%%%%%%%%%%%%%%%%%%%%%%%%%%%%%%%%%%%%%%%%%%%%%%%%%%%%%%%%%%%%%%
\begin{proof}
Write $V_{\be(k)}=\sgn(V_{\be(k)}) |V_{\be(k)}|$.
By the Schwarz inequality, Lemmas \ref{prop of v} and \ref{bound tb}, we have
\begin{align*}
&  |\inner{\Psi}{L_{21}(k)\Phi}| \\
& \leq \int_1^\infty \norm{\sgn(V_{\be(k)})|V_{\be(k)}|^{\frac{1}{2}} R_{t^2}\Psi}
  \norm{|V_{\be(k)}|^{\frac{1}{2}} R_{t^2}(\bk)(e^{-i\bk\cdot\bx}-1)\braket{\bx}^{-1}\Phi} t^2dt \notag \\
& \leq \left(\int_0^\infty \norm{T_\bA^{\frac{1}{4}} R_{t^2}\Psi}^2 t^2dt \right)^{\frac{1}{2}}
  \left(\int_0^\infty \norm{T_{\bA-\bk}^{\frac{1}{4}} R_{t^2}(\bk) (e^{-i\bk\cdot\bx}-1)
  \braket{\bx}^{-1}\Phi}^2 t^2dt \right)^{\frac{1}{2}} \notag \\
& = \left( \frac{\pi}{4} \norm{\Psi}^2\right)^{\frac{1}{2}}
    \left( \frac{\pi}{4} \norm{(e^{-i\bk\cdot\bx}-1)\braket{\bx}^{-1}\Phi}^2\right)^{\frac{1}{2}},
\end{align*}
where we used  $\int_0^\infty at^2/(a^2+t^2)^2 dt = \pi/4$ for $a>0$.
Since 
 $|(e^{-i\bk\cdot\bx}-1)\braket{\bx}^{-1}|\leq |\bk|$ and $\pi/4<1$, 
\eqref{xx271} follows.
\end{proof}
Bounds for $L_{22}(k)$ and $L_{23}(k)$ are given in the following.
%%%%%%%%%%%%%%%%%%%%%%%%%%%%%%%%%%%%%%%%%%%%%%%%%%%%%%%%%%%%%%%%
\begin{lem}{\label{bound l22}}
For $k=(\bk,j)\in \RR^3\times\{1,2\}$, we have 
\begin{align*}
&  \norm{L_{22}(k)} \leq 2|\bk|, \\
&  \norm{L_{23}(k)} \leq |\bk|^2. 
\end{align*}
\end{lem}
%%%%%%%%%%%%%%%%%%%%%%%%%%%%%%%%%%%%%%%%%%%%%%%%%%%%%%%%%%%%%%%%
\begin{proof}
We have
\begin{align*}
 \norm{L_{22}(k)}
 \leq 2 \int_1^\infty \norm{t^2 R_{t^2}} \norm{V_{\be(k)}R_{t^2}^{\frac{1}{2}}}
 \norm{R_{t^2}^{\frac{1}{2}} V_\bk} \norm{R_{t^2}(\bk) \braket{\bx}^{-1}} dt.
\end{align*}
By Lemma \ref{bound tb}, we have $\norm{V_{\be(k)}R_{t^2}^{\frac{1}{2}}} \leq 1$ and 
$\norm{R_{t^2}^{\frac{1}{2}} V_\bk} = \norm{V_\bk R_{t^2}^{\frac{1}{2}}} \leq |\bk|$.
Thus
\begin{align*}
  \norm{L_{22}(k)} \leq 2 \int_1^\infty  |\bk|\cdot t^{-2} dt = 2|\bk|.
\end{align*}
Similarly, we have
\begin{align*}
 \norm{L_{23}(k)}
 \leq |\bk|^2 \int_1^\infty \norm{t^2 R_{t^2}} \norm{V_{\be(k)}R_{t^2}}
  \norm{R_{t^2}(\bk)\braket{\bx}^{-1}} dt 
   \leq |\bk|^2 \int_1^\infty t^{-3} dt  \leq |\bk|^2.
\end{align*}
\end{proof}
%%%%%%%%%%%%%%%%%%%%%%%%%%%%%%%%%%%%%%%%%%%%%%%%%%%%%%%%%%%%%%%%
\subsection{Estimate on $L_3$}
We shall estimate $L_3(k)$.
A crucial  property of $L_3(k)$ is the identity
\begin{align*}
  L_3(k) = \frac{i\pi}{4} [T_\bA^{\frac{1}{2}}, \be(k)\cdot \bx] = \frac{i\pi}{4} [H_m-E_m,\be(k)\cdot \bx],
\end{align*}
which will enable us to obtain an infrared regular bound for $L_3(k)$.
This was due to  \cite[p.214]{hir03} and \cite[(7.7)]{hhs05}.
For operators $A$ and $B$, we define the quadratic form $[A,B]_\mathrm{w}$ as
\begin{align*}
  [A,B]_\mathrm{w}(u,v) = \inner{Au}{Bv}-\inner{Bu}{Av}, \qquad u,v\in \dom(A)\cap \dom(B).
\end{align*}
We also write this as $\inner{u}{[A,B]_\mathrm{w}v}$.
%%%%%%%%%%%%%%%%%%%%%%%%%%%%%%%%%%%%%%%%%%%%%%%%%%%%%%%%%%%%%%%%
\begin{lem}{\label{idtty l3}}
For $\Psi\in \Hfin$ and $\Phi\in \dom(H_m)\cap \dom(|\bx|)$,
\begin{align*}
  \inner{\Psi}{L_3(k)\Phi}
  = \frac{i\pi}{4} \inner{\Psi}{[H_m-E_m, \be(k)\cdot\bx]_\mathrm{w}\Phi}.
\end{align*}
In particular, $\be(k)\cdot\bx\Phi_m \in \dom(H_m)$ and it holds that 
\begin{align*}
  L_3(k)\Phi_m  = \frac{i\pi}{4} (H_m-E_m)(\be(k)\cdot\bx)\Phi_m. 
\end{align*}
\end{lem}
%%%%%%%%%%%%%%%%%%%%%%%%%%%%%%%%%%%%%%%%%%%%%%%%%%%%%%%%%%%%%%%%
\begin{proof}
By the definition of $L_3$ we have
\begin{align*}
  \inner{\Psi}{L_3(k)\Phi}
  = \int_0^\infty \inner{R_{t^2}\Psi}{V_{\be(k)}R_{t^2}\Phi} t^2 dt. 
\end{align*}
We note that, by Lemma \ref{diamag}, $R_{t^2}\Psi, R_{t^2}\Phi \in \dom(|\bx|)$ for $t>0$.
Since $T_\bA R_{t^2} = \one -t^2 R_{t^2}$, we have $T_\bA R_{t^2}\Psi, T_\bA R_{t^2} \Phi\in \dom(|\bx|)$.
%By Proposition \ref{bd prnp} and $\Psi,\Phi\in \dom(N)$, we know that $R_{t^2}\Psi, R_{t^2}\Phi\in \dom(|\bp|)$.
For any $\psi\in\Hfin$, we have  $V_{\be(k)}\psi = \frac{i}{2} [T_\bA, \be(k)\cdot\bx]\psi$.
Thus  for $\varphi\in \dom( (\be(k)\cdot \bx) T_\bA)$, it follows that 
\begin{align}
  \inner{\psi}{V_{\be(k)}\varphi}
  = \frac{i}{2} \big( \inner{T_\bA \psi}{\be\cdot\bx\varphi} - \inner{\psi}{(\be\cdot\bx)T_\bA 
   \varphi} \big).
  \label{xy285}
\end{align}
Since $\Hfin$ is a core for $T_\bA$, \eqref{xy285} can be extended for all $\psi\in\dom(T_\bA) \cap \dom(|\bx|)$.
Hence we have
\begin{align*}
  \inner{R_{t^2}\Psi}{V_{\be(k)}R_{t^2}\Phi}
  &= \frac{i}{2} \big( \inner{T_\bA R_{t^2}\Psi}{ (\be\cdot\bx) R_{t^2}\Phi}
  - \inner{R_{t^2}\Psi}{(\be\cdot\bx)T_\bA R_{t^2}\Phi} \big) \\
  &= \frac{i}{2} \big( \inner{\be\cdot\bx\Psi}{R_{t^2}\Phi} - \inner{R_{t^2}\Psi}{\be\cdot\bx\Phi} \big) \\
  &= \frac{i}{2t^2} \big( -\inner{\be\cdot\bx\Psi}{T_\bA R_{t^2}\Phi} 
     + \inner{T_\bA R_{t^2}\Psi}{\be\cdot\bx\Phi} \big).
\end{align*}
By the formula \eqref{sqrt int}, we have
\begin{align*}
  \inner{\Psi}{L_3(k)\Phi}
  = \frac{i\pi}{4} \inner{\Psi}{[T_\bA^{\frac{1}{2}},\be\cdot\bx]_\mathrm{w} \Phi}
  = \frac{i\pi}{4} \inner{\Psi}{[H_m-E_m,\be\cdot\bx]_\mathrm{w} \Phi}.
\end{align*}
\end{proof}
%%%%%%%%%%%%%%%%%%%%%%%%%%%%%%%%%%%%%%%%%%%%%%%%%%%%%%%%%%%%%%%%
\subsection{Proof of Proposition \ref{pdenb}}
\begin{proof}[Proof of Proposition \ref{pdenb}:]
  By the singular and non-local pull-through formula \eqref{puth} and the decomposition \eqref{xy243}, we have
\begin{align*}
 \norm{a(k)\Phi_m}
 \leq& |\phiome(\bk)| \frac{2\sqrt{2}}{\pi}
 \Big( \ome_m(\bk)^{-1}\norm{L_1(k)} \norm{\braket{\bx}^2\Phi_m}
      + \ome_m(\bk)^{-1}\norm{L_2(k)} \norm{\braket{\bx}\Phi_m} \\
     &+ \norm{(H_m-E_m+\ome_m(\bk))^{-1}L_3(k)\Phi_m}
   \Big),
\end{align*}
where we used the inequality $\norm{(H_m-E_m+\ome_m(\bk))^{-1}}\leq \ome_m(\bk)^{-1}$.
By Lemmas \ref{bound l1}, \ref{bound l21}, \ref{bound l22} and \eqref{xy266}, we have
\begin{align}
  \norm{L_1(k)}\leq 2|\bk|, \qquad
  \norm{L_2(k)}\leq |\bk|+2|\bk|+|\bk|^2  \label{xy293}
\end{align}
Moreover, by Lemma \ref{idtty l3}, we have
\begin{align}
& \norm{(H_m-E_m+\ome_m(\bk))^{-1} L_3(k)\Phi_m} \notag \\
&\leq \frac{\pi}{4} \norm{(H_m-E_m+\ome_m(\bk))^{-1}(H_m-E_m)(\be(k)\cdot\bx)\Phi_m}  
%&\leq \frac{\pi}{4} \norm{(H_m-E_m+\ome_m(\bk))^{-1}(H_m-E_m)} \norm{\be(k)\cdot\bx\Phi_m} \notag \\
\leq   \frac{\pi}{4} \norm{|\bx|\Phi_m}.   \label{xy294}
\end{align}
By assumption (A4), the bounds
\begin{align}
  \sup_{0<m< m_0} \norm{|\bx|\Phi_m} <\infty , \qquad
  \sup_{0<m< m_0}\norm{\braket{\bx}^2\Phi_m} <\infty  \label{xy295}
\end{align}
hold.
Therefore, by \eqref{xy293}--\eqref{xy295}, we have
\begin{align*}
  \norm{a(k)\Phi_m}
  \leq
  C |\phiome(\bk)| \bigg( \frac{|\bk|+|\bk|^2}{\ome_m(\bk)}+1 \bigg)
  \leq C \frac{|\phih(\bk)|}{\ome(\bk)^{\frac{1}{2}}} (2+|\bk|),
  \qquad 0<m< m_0,
\end{align*}
for some $C>0$.
This immediately implies \eqref{xy237}.
The integrability of $\norm{a(k)\Phi_m}^2$ follows from the assumption (A2).
\end{proof}
%%%%%%%%%%%%%%%%%%%%%%%%%%%%%%%%%%%%%%%%%%%%%%%%%%%%%%%%%%%%%%%%
\section{Equicontinuity and spatial localization of photon} \label{sec:slp}
In this section, we show that the photon of the massive ground state $\Phi_m$ are
spatially localized uniformly in $0<m<m_0$.
Throughout this section, we assume (A1)--(A4).
%%%%%%%%%%%%%%%%%%%%%%%%%%%%%%%%%%%%%%%%%%%%%%%%%%%%%%%%%%%%%%%%
\subsection{Continuity of $J(k)$}
%It is sufficient to show Lemma \ref{sp bd 2} to prove Lemma \ref{photonloc2}.  
%We will prove Proposition \ref{photonloc} by showing (b) in Lemma \ref{sp bd 2}.
%We need to estimate the difference $a(k)\Phi_m-a(k')\Phi_m$.
We shall show the continuity of $k\mapsto J(k)$ in this section. 
We decompose $J(k)-J(k')$ as follows
\begin{align*}
  J(k)-J(k')= \Delta J_1 + \Delta J_2,
\end{align*}
with
\begin{align*}
 \Delta J_1 & = \int_0^\infty R_{t^2}(V_{\be(k)}-V_{\be(k')}) e^{-i\bk\cdot\bx} R_{t^2} t^2 dt, \\
 \Delta J_2 & = \int_0^\infty R_{t^2} V_{\be(k')} ( e^{-i\bk\cdot\bx}-e^{-i\bk'\cdot \bx} ) R_{t^2} t^2 dt.
\end{align*}
%%%%%%%%%%%%%%%%%%%%%%%%%%%%%%%%%%%%%%%%%%%%%%%%%%%%%%%%%%%%%%%%
\begin{lem}{\label{bd DeltaJ1}}
Let $k=(\bk,j)$ and $k'=(\bk',j)$. For any $\Phi\in \dom(|\bx|^{\frac{1}{2}})$ 
it follows that 
\begin{align}
  \norm{\Delta J_1 \Phi} \leq 
  |\be(k)-\be(k')|\Big(\norm{\Phi}+|\bk|^{\frac{1}{2}} \norm{|\bx|^{\frac{1}{2}}\Phi}\Big).  \label{zz622}
\end{align}
\end{lem}
%%%%%%%%%%%%%%%%%%%%%%%%%%%%%%%%%%%%%%%%%%%%%%%%%%%%%%%%%%%%%%%%
\begin{proof}
Set $\be=\be(k)$ and $\be'=\be(k')$. Since 
\begin{align*}
 V_\be-V_{\be'}  = (\be-\be')\cdot(\bp-\bA(\bx))=V_{\be-\be'} 
 = \sgn(V_{\be-\be'}) |V_{\be-\be'}|,
\end{align*}
for any $\Psi \in \sH$, we have
\begin{align}
 \big|\inner{\Psi}{\Delta J_1\Phi}\big|
 \leq 
 \int_0^\infty \norm{|V_{\be-\be'}|^{\frac{1}{2}}R_{t^2}\Psi}
 \norm{|V_{\be-\be'}|^{\frac{1}{2}} e^{-i\bk\cdot\bx} R_{t^2}\Phi} t^2 dt \notag \\
 \leq 
 |\be-\be'| \int_0^\infty \norm{T_\bA^{\frac{1}{4}}R_{t^2}\Psi}
 \norm{T_\bA^{\frac{1}{4}} e^{-i\bk\cdot\bx} R_{t^2}\Phi} t^2 dt, \label{zz626}
\end{align}
where we used Lemma \ref{prop of v}. We note that 
\begin{align*}
&\norm{T_\bA^{\frac{1}{4}} e^{-i\bk\cdot\bx} R_{t^2}\Phi}^2 
= \norm{T_{\bA+\bk}^{\frac{1}{4}} R_{t^2}\Phi}^2
= \inner{R_{t^2}\Phi}{|\bp-\bA(\bx)-\bk|R_{t^2}\Phi} \\
&\leq \inner{R_{t^2}\Phi}{|\bp-\bA(\bx)| R_{t^2}\Phi}
  + |\bk| \inner{R_{t^2}\Phi}{R_{t^2}\Phi}
  = \norm{T_\bA^{\frac{1}{4}} R_{t^2}\Phi}^2 + |\bk| \norm{R_{t^2}\Phi}^2.
\end{align*}
Thus  \eqref{zz626} is bounded by
\begin{align*}
&  |\be-\be'|\bigg(\int_0^\infty \norm{T_\bA^{\frac{1}{4}}R_{t^2}\Psi}^2 t^2 dt \bigg)^{\frac{1}{2}}
  \bigg(\int_0^\infty \big( \norm{T_\bA^{\frac{1}{4}} R_{t^2}\Phi}^2 + |\bk|\norm{R_{t^2}\Phi}^2\big) t^2 dt \Bigg)^{\frac{1}{2}}\\
&=  |\be-\be'| \Big( \frac{\pi}{4}\norm{\Psi}^2\Big)^{\frac{1}{2}}
   \Big( \frac{\pi}{4}\norm{\Phi}^2 + \frac{\pi}{4}|\bk| \norm{T_\bA^{-\frac{1}{4}}\Phi}^2 \Big)^{\frac{1}{2}}.
\end{align*}
From the diamagnetic inequality and Hardy-Kato's inequality we have
\begin{align}\label{HK2}
  \norm{T_\bA^{-\frac{1}{4}}\Phi}^2 \leq \norm{|\bp|^{-\frac{1}{2}}\Phi}^2 \leq \frac{\pi}{2} \norm{|\bx|^{\frac{1}{2}}\Phi}^2.
\end{align}
Therefore we have the bound
\begin{align*}
  \norm{\Delta J_1\Phi} 
&= \sup_{\norm{\Psi}=1} |\inner{\Psi}{\Delta J_1 \Phi}| 
\leq  \frac{\pi}{4} |\be-\be'|  \big(\norm{\Phi}^2 + \frac{\pi}{2}|\bk|\norm{|\bx|^{\frac{1}{2}}\Phi}^2 \big)^{\frac{1}{2}} \\
&\leq  |\be-\be'|  (\norm{\Phi} + |\bk|^{\frac{1}{2}} \norm{|\bx|^{\frac{1}{2}}\Phi}),
\end{align*}
which implies \eqref{zz622}.
\end{proof}
We decompose $\Delta J_2$ into two terms:
\begin{align*}
  \Delta J_2 = \Delta J_{21} + \Delta J_{22},
\end{align*}
with
\begin{align*}
\Delta J_{21} &= \int_0^1 R_{t^2} V_{\be(k')} ( e^{-i\bk\cdot\bx}-e^{-i\bk'\cdot \bx} ) R_{t^2} t^2 dt, \\
\Delta J_{22} &= \int_1^\infty R_{t^2} V_{\be(k')} ( e^{-i\bk\cdot\bx}-e^{-i\bk'\cdot \bx} ) R_{t^2} t^2 dt.
\end{align*}
%%%%%%%%%%%%%%%%%%%%%%%%%%%%%%%%%%%%%%%%%%%%%%%%%%%%%%%%%%%%%%%%
\begin{lem}{\label{bd DeltaJ21}}
 For any $\Phi\in \dom(|\bx|^2)$,
\begin{align*}
   \norm{\Delta J_{21}\Phi} \leq 2|\bk-\bk'|  \norm{\braket{\bx}^2\Phi}.
\end{align*}
\end{lem}
%%%%%%%%%%%%%%%%%%%%%%%%%%%%%%%%%%%%%%%%%%%%%%%%%%%%%%%%%%%%%%%%
\begin{proof}
  The proof is similar to that  of Lemma \ref{bound l1}.
\end{proof}
%%%%%%%%%%%%%%%%%%%%%%%%%%%%%%%%%%%%%%%%%%%%%%%%%%%%%%%%%%%%%%%%
\begin{lem}{\label{bd DeltaJ22}}
Let $k=(\bk,j)$ and $k'=(\bk',j)$. For any $\Phi\in \dom(|\bx|^{\frac{1}{2}})$ 
it holds that 
\begin{align}
 \norm{\Delta J_{22} \Phi} \leq 
 2|\bk-\bk'|(1+|\bk'|) \norm{\Phi}
 + ||\bk'|^2-|\bk|^2| \norm{\Phi}
 + |\bk-\bk'| \norm{|\bx|\Phi}. \label{zz638}
\end{align}  
\end{lem}
%%%%%%%%%%%%%%%%%%%%%%%%%%%%%%%%%%%%%%%%%%%%%%%%%%%%%%%%%%%%%%%%
\begin{proof}
Recall that $R_{t^2}(\bk)= e^{-i\bk\cdot\bx}R_{t^2}e^{i\bk\cdot\bx}=((\bp-\bA(\bx)+\bk)^2+t^2)^{-1}$. 
Then we have
\begin{align*}
& ( e^{-i\bk\cdot\bx}-e^{-i\bk'\cdot \bx} ) R_{t^2} 
= ( R_{t^2}(\bk) - R_{t^2}(\bk')) e^{-i\bk\cdot\bx} 
  + R_{t^2}(\bk') (e^{-i\bk\cdot\bx}-e^{-i\bk'\cdot \bx}) \\
&= R_{t^2}(\bk')(T_{\bA-\bk'} -T_{\bA-\bk}) R_{t^2}(\bk)e^{-i\bk\cdot\bx}
  + R_{t^2}(\bk') (e^{-i\bk\cdot\bx}-e^{-i\bk'\cdot \bx}) \\
&= 2R_{t^2}(\bk') V_{\bk'-\bk} R_{t^2}(\bk)e^{-i\bk\cdot\bx}
  + (|\bk'|^2-|\bk|^2) R_{t^2}(\bk') R_{t^2}(\bk)e^{-i\bk\cdot\bx} \\
&\hspace{3cm}   + R_{t^2}(\bk') (e^{-i\bk\cdot\bx}-e^{-i\bk'\cdot \bx}).
\end{align*}
According to this decomposition, $\Delta J_{22}$ can be furthermore decomposed into three terms:
\begin{align}
 \Delta J_{22} = \Delta J_{221} + \Delta J_{222} + \Delta J_{223} \label{zz639}
\end{align}
with 
\begin{align*}
 \Delta J_{221} 
&= \int_1^\infty R_{t^2} V_{\be(k')} 2R_{t^2}(\bk') V_{\bk'-\bk} R_{t^2}(\bk)e^{-i\bk\cdot\bx}t^2 dt , \\
 \Delta J_{222} 
&= \int_1^\infty R_{t^2} V_{\be(k')} (|\bk'|^2-|\bk|^2) R_{t^2}(\bk') R_{t^2}(\bk) e^{-i\bk\cdot\bx} t^2 dt, \\
 \Delta J_{223} 
&= \int_1^\infty R_{t^2} V_{\be(k')} R_{t^2}(\bk') (e^{-i\bk\cdot\bx}-e^{-i\bk'\cdot \bx}) t^2 dt.
\end{align*}
We can estimate $\Delta J_{221}$ as follows.
\begin{align}
 \norm{\Delta J_{221}\Phi}
 &\leq 
 2 \int_1^\infty \norm{t^2R_{t^2}} \norm{V_{\be(k')} R_{t^2}(\bk')^{\frac{1}{2}}}
   \norm{ R_{t^2}(\bk')^{\frac{1}{2}}V_{\bk'-\bk}} \norm{R_{t^2}(\bk)e^{-i\bk\cdot\bx}\Phi} dt  \notag \\
 & \leq
 2 \int_1^\infty |\bk'-\bk|(1+|\bk'|) t^{-2} \norm{\Phi} dt  = 2 |\bk'-\bk|(1+|\bk'|) \norm{\Phi},  \label{zz643}
\end{align}
where we used bounds below: 
\begin{align*}
\norm{t^2R_{t^2}} &\leq 1,\\
\norm{R_{t^2}(\bk)e^{-i\bk\cdot\bx}\Phi} &\leq t^{-2}\norm{\Phi},\\
 \norm{V_{\be(k')} R_{t^2}(\bk')^{\frac{1}{2}}} 
& = \norm{V_{\be(k')} e^{-i\bk'\cdot\bx} R_{t^2}^{\frac{1}{2}} e^{i\bk\cdot\bx}} 
 = \norm{V_{\be(k')} R_{t^2}^{\frac{1}{2}}} \leq 1, \notag \\
 \norm{ R_{t^2}(\bk')^{\frac{1}{2}} V_{\bk'-\bk}} 
& = \norm{ V_{\bk'-\bk} R_{t^2}(\bk')^{\frac{1}{2}} }
 = \norm{ (\bk'-\bk)\cdot(\bp-\bA(\bx)) e^{-i\bk'\cdot\bx} R_{t^2}^{\frac{1}{2}} e^{i\bk'\cdot\bx}}  \notag \\
& = \norm{ (\bk'-\bk)\cdot(\bp-\bA(\bx)-\bk') R_{t^2}^{\frac{1}{2}}}  
 \leq |\bk'-\bk| \big( \norm{|\bp-\bA(\bx)|R_{t^2}^{\frac{1}{2}}} + |\bk'|\big) \notag \\
&  \leq |\bk'-\bk| \big( 1 + |\bk'| \big).
\end{align*}
Next we estimate $\Delta J_{222}$ as 
\begin{align}
 \norm{\Delta J_{222}\Phi} 
\leq 
 ||\bk'|^2-|\bk|^2|\int_1^\infty \norm{t^2R_{t^2}} \norm{V_{\be(k')}R_{t^2}(\bk')}
 \norm{R_{t^2}(\bk)} \norm{\Phi} dt 
 \leq ||\bk'|^2-|\bk|^2| \norm{\Phi}. \label{zz648}
\end{align}
Finally we estimate $\Delta J_{223}$. We see that 
\begin{align}
 \norm{\Delta J_{223}\Phi}
& = \sup_{\norm{\Psi}=1} |\inner{\Psi}{\Delta J_{223} \Phi}|  \notag \\
& \leq \sup_{\norm{\Psi}=1} \int_1^\infty \norm{|V_{\be'}|^{\frac{1}{2}}R_{t^2}\Psi}
   \norm{|V_{\be'}|^{\frac{1}{2}} R_{t^2}(\bk')(e^{-i\bk\cdot\bx}-e^{-i\bk'\cdot\bx})\Phi} t^2dt \notag \\
& \leq \sup_{\norm{\Psi}=1} \int_1^\infty \norm{T_\bA^{\frac{1}{4}}R_{t^2}\Psi}
   \norm{|V_{\be'}|^{\frac{1}{2}} R_{t^2}(e^{-i(\bk-\bk')\cdot\bx}-1)\Phi} t^2 dt \notag \\
& \leq \sup_{\norm{\Psi}=1} \int_1^\infty \norm{T_\bA^{\frac{1}{4}}R_{t^2}\Psi}
   \norm{T_\bA^{\frac{1}{4}} R_{t^2}(e^{-i(\bk-\bk')\cdot\bx}-1)\Phi} t^2 dt \notag \\
& \leq \sup_{\norm{\Psi}=1} \Big(\int_0^\infty \! \norm{T_\bA^{\frac{1}{4}}R_{t^2}\Psi}^2 t^2 dt \Big)^{\frac{1}{2}}
  \Big(\int_0^\infty \! \norm{T_\bA^{\frac{1}{4}} R_{t^2}(e^{-i(\bk-\bk')\cdot\bx}-1)\Phi}^2 t^2dt \Big)^{\frac{1}{2}} \notag \\
& = \frac{\pi}{4}\norm{(e^{-i(\bk-\bk')\cdot\bx}-1)\Phi}  \leq  |\bk-\bk'|  \norm{|\bx|\Phi}. \label{zz655}
\end{align}
Combining estimates \eqref{zz643}, \eqref{zz648} and \eqref{zz655}, we get 
\eqref{zz622}.
\end{proof}
%%%%%%%%%%%%%%%%%%%%%%%%%%%%%%%%%%%%%%%%%%%%%%%%%%%%%%%%%%%%%%%%
\begin{lem}{\label{akDiff}}
 For almost every $k,k'\in \RR^3\times\{1,2\}$, it follows that 
 \begin{align*} 
 \sup_{0<m<m_0}\norm{(J(k)-J(k'))\Phi_m}
& \leq  |\be(k)-\be(k')|(1+|\bk|^{\frac{1}{2}} D) + 2D|\bk-\bk'| \\
& \quad + 2|\bk-\bk'|(1+|\bk'|) 
  + ||\bk'|^2-|\bk|^2|  + |\bk-\bk'| D, 
\end{align*}
where $D$ is a constant defined by
$
  D = \sup_{0<m<m_0} \norm{\braket{\bx}^2\Phi_m}$.
\end{lem}
%%%%%%%%%%%%%%%%%%%%%%%%%%%%%%%%%%%%%%%%%%%%%%%%%%%%%%%%%%%%%%%%
\begin{proof}
  This is a consequence of Lemmas \ref{bd DeltaJ1}, \ref{bd DeltaJ21} and 
\ref{bd DeltaJ22}.
\end{proof}
%%%%%%%%%%%%%%%%%%%%%%%%%%%%%%%%%%%%%%%%%%%%%%%%%%%%%%%%%%%%%%%%

\subsection{Equicontinuity of $\{a(k)\Phi_m\}$}
In this section we shall show that $\{a(k)\Phi_m\}_{0<m<m_0}$ is equicontinuous. In order to investigate more general setting 
on equicontinuity we introduce domain $D_\ep$. 
For any $0<\ep\ll 1$, we define a measurable set $D_\ep \subset \RR^3$ so that
for any $\rho\in L^2(\RR^3)$, 
\begin{align*}
  \lim_{\ep\to+0} \int_{D_\ep} |\rho(\bk)|^2 d\bk = 0.
\end{align*}
%\end{df}
%%%%%%%%%%%%%%%%%%%%%%%%%%%%%%%%%%%%%%%%%%%%%%%%%%%%%%%%%%%%%%%%
\begin{exa}{\label{RemOfD}}
 An example of $D_\ep$ is given by
\begin{align}\label{DE}
  D_\ep = \{\bk \in \RR^3 \mid k_1^2+k_2^2 \leq \ep\} \cup \{\bk\in\RR^3 \mid |\bk| \geq 1/\ep\}
\end{align}
For simplicity, the set $\{ k=(\bk,j) \mid \bk\in D_\ep, j=1,2\}$ is also denoted by $D_\ep$.
\end{exa}
\begin{thm}[Equicontinuity]
\label{econt}
Suppose (A1)--(A4). Then 
\begin{align}
  \sup_{0<m<m_0}  \int_{D_\ep^\mathrm{c}}\norm{a(k)\Phi_m-a(k-s)\Phi_m}^2 dk 
 \to 0 \quad (|\bs|\to 0),   \label{condB}
\end{align}
where $D_\ep$ is given by \eqref{DE}.
\end{thm}
\begin{proof}
We fix $\ep>0$ arbitrarily.
Note that  $D_\ep$ satisfy 
\begin{itemize}
\item[\rm (d1)] $D_\ep \subset D_{\ep'}$ for $\ep<\ep'$,
\item[\rm (d2)] $\mathrm{dist}(D_\ep^\mathrm{c}, D_{\frac{\ep}{2}})\geq \frac{\ep}{2}$.
\end{itemize}
By the definition, $\be(\bk,j), j=1,2$ are uniformly continuous in $D_\ep^\mathrm{c}$.
For $k=(\bk,j)\in D_\ep^\mathrm{c}$, we set $k'= (\bk-\bs,j)$.
By (d2), $|\bs|<\frac{\ep}{2}$ implies $k' \in D_{\frac{\ep}{2}}^{\mathrm{c}}$,
and hence $\ome(\bk), \ome(\bk')\geq \frac{\ep}{2}$.
We decompose $a(k)\Phi_m-a(k')\Phi_m$ into three terms:
\begin{align*}
  a(k)\Phi_m - a(k')\Phi_m = A_1 +  A_2+  A_3,
\end{align*}
where
\begin{align*}
&  A_1 = \phiome(\bk)(H_m-E_m+\ome_m(\bk))^{-1}(J(k)-J(k'))\Phi_m, \\
&  A_2 = \phiome(\bk) 
         \big\{ (H_m-E_m+\ome_m(\bk))^{-1}- (H_m-E_m+\ome_m(\bk'))^{-1} \big\}
         J(k')\Phi_m, \\
&  A_3 = (\phiome(\bk)-\phiome(\bk'))(H_m-E_m+\ome_m(\bk'))^{-1}J(k')\Phi_m.
\end{align*}
By  Lemma \ref{akDiff}, we can estimate the norm of $A_1$ as follows:
\begin{align*}
  \norm{A_1}
&\leq |\phiome(\bk)| \ome_m(\bk)^{-1} \norm{(J(k)-J(k'))\Phi_m} 
\leq |\phiome(\bk)| \frac{2}{\ep} \norm{(J(k)-J(k'))\Phi_m} \\
&\leq C|\phiome(\bk)|  \Big( |\be(\bk,j)-\be(\bk-\bs,j)| + |\bs|\Big),
\end{align*}
where $C$ is a constant independent of $\bk,\bs$ and $m$. 
Thus we have
\begin{align}
  \lim_{|\bs|\to 0} \int_{D_\ep^\mathrm{c}} \norm{A_1}^2 dk = 0.   \label{669}
\end{align}
Next we consider $A_2$. By Corollary  \ref{def of j}, we have
\begin{align*}
  \norm{A_2} 
&\leq |\phiome(\bk)| \ome_m(\bk)^{-1}\ome_m(\bk')^{-1}|\ome_m(\bk)-\ome_m(\bk')|\norm{J(k')} \leq |\phiome(\bk)| \frac{4}{\ep^2} |\bk-\bk'|  \frac{1}{\sqrt{2}} \\
& =   \frac{2\sqrt{2}}{\ep^2}|\phiome(\bk)|  |\bs|.
\end{align*}
Thus  we have
\begin{align}
  \lim_{|\bs|\to 0} \int_{D_\ep^\mathrm{c}} \norm{A_2}^2 dk = 0. \label{673}
\end{align}
The norm of $A_3$ can be similarly estimated as follows.
\begin{align*}
  \norm{A_3}
\leq |\phiome(\bk)-\phiome(\bk-\bs)| \frac{\sqrt{2}}{\ep}.
\end{align*}
Since $\phiome \in L^2(\RR^3_\bk)$, the shift  $\bs\mapsto \phiome(\cdot-\bs)$ is 
strongly continuous, and hence
\begin{align}
  \lim_{|\bs|\to 0} \int_{D_\ep^\mathrm{c}} \norm{A_3}^2 dk = 0. \label{675}
\end{align}
Therefore, by \eqref{669}, \eqref{673} and \eqref{675}, 
we can show \eqref{condB}. 
\end{proof}

\subsection{Spatial localization of photon}
%We modify a method developed by C. G\'erard \cite{ge00,ge06}.
Let $\cB(K)$ be the set of bounded operator on $K$. For $T\in \cB(W)$ with $\norm{T}\leq 1$, we define the second quantization of $T$, 
$\Gamma(T)\in \mathcal{B}(\sF)$,  by
\begin{align*}
  \Gamma(T)= \oplus_{n=0}^\infty (\oplus^n T).
\end{align*}
Here we set $\oplus^0T=\one$. 
Let $j\in C_0^\infty([0,\infty))$ be a function such that $0\leq j(s)\leq 1$ and 
\begin{align*}
  j(s) =
  \begin{cases}
    1 &  0 \leq s \leq 1, \\
    0 &  s \geq 2.
  \end{cases}
\end{align*}
For $R>0$, we set $\chi(\by)=j(|\by|)$ and $\chi_R= \chi(i\nabla_\bk/R)$ and $\Gamma_R
= \Gamma(\chi_R) = \one_W \tensor \Gamma(\chi_R) $.
In this section we shall prove the proposition below: 
\begin{prop}[Spatial localization of photon]{\label{photonloc}}
  Assume (A1)--(A4). Then it holds that 
\begin{align}
    \lim_{R\to\infty} \sup_{0<m<m_0}\norm{(\one-\Gamma_R) \Phi_m} = 0. \label{sploc}
\end{align}
\end{prop}
The proof of Proposition \ref{photonloc} is given after general lemmas stated below. 
For $f\in L^2(\RR^3)$, it holds that
\begin{align}
 \chi_R f = (2\pi)^{-\frac{3}{2}}\int_{\RR^3} \hat\chi(\bs) f(\cdot -R^{-1}\bs) d\bs. \label{repr chiR}
\end{align}
%which is a strong Riemann integral. 
Note that $\hat\chi$ is a rapidly decreasing smooth function.
We can extend this type formula to the state in $\sH$.
%%%%%%%%%%%%%%%%%%%%%%%%%%%%%%%%%%%%%%%%%%%%%%%%%%%%%%%%%%%%%%%%
\begin{lem}{\label{lem of repr dGw}}
  For $\Phi\in \dom(\Nf^{\frac{1}{2}})$, we have 
  \begin{align}
 \norm{d\Gamma(\chi_R)^{\frac{1}{2}}\Phi}^2
 = (2\pi)^{-\frac{3}{2}}\int_{\RR^3}d\bs\int  \hat\chi(\bs)
   \inner{a(k)\Phi}{a( k-R^{-1}s )\Phi}  dk,   \label{repr dgw}
\end{align}
where $k-R^{-1}s = (\bk-R^{-1}\bs, j)$ with $k=(\bk,j)\in\RR^3\times\{1,2\}$,
and the integral \eqref{repr dgw} is absolutely convergent.
\end{lem}
%%%%%%%%%%%%%%%%%%%%%%%%%%%%%%%%%%%%%%%%%%%%%%%%%%%%%%%%%%%%%%%%
\begin{proof}
The particle part is irrelevant to this result, so for simplicity, we only consider the field part.
For each $n$-particle part $\Phi^{(n)}$, from \eqref{repr chiR}, we have
\begin{align*}
 (\chi_R\tensor \one_{\stensor^{n-1}W}) \Phi^{(n)}(k_1,\dots,k_n) = 
 (2\pi)^{-\frac{3}{2}}\int_{\RR^3} \hat\chi(\bs) \Phi^{(n)} (k_1-R^{-1}s, k_2, \ldots, k_n) d\bs,
\end{align*}
which is a strong integral in $\stensor^n W$.
Thus  by the symmetry of the state and the definition of $a(k)$, we have
\begin{align*}
  ( \chi_R^{(n)} \Phi^{(n)})(k_1,\ldots,k_n) 
& = n(2\pi)^{-\frac{3}{2}}\int_{\RR^3} \hat\chi(\bs) \Phi^{(n)}(k_1-R^{-1} s, k_2,\ldots,k_n) d\bs  \\
& = \sqrt{n} (2\pi)^{-\frac{3}{2}}\int_{\RR^3} \hat\chi(\bs) (a(k_1-R^{-1}s)\Phi)^{(n-1)}(k_2,\ldots,k_n) d\bs.
\end{align*}
Since $\Phi^{(n)}(k,\cdot)=n^{-\frac{1}{2}}(a(k)\Phi)^{(n-1)}(\cdot)$, we have
\begin{align*}
  \biginner{\Phi^{(n)}}{\chi_R^{(n)}\Phi^{(n)}}
 = (2\pi)^{-\frac{3}{2}}\int_{\RR^3}d\bs \int   \hat\chi(\bs)  
   \inner{(a(k)\Phi)^{(n-1)}}{(a( k-R^{-1}s )\Phi)^{(n-1)}}_{\stensor^{n-1} W} dk,
\end{align*}
for $n=1,2,\ldots$, 
and 
\begin{align*}
 \sum_{n=1}^\infty \int_{\RR^3}d\bs\int   |\hat\chi(\bs)| 
 \big|\inner{(a(k)\Phi)^{(n-1)}}{(a(k-R^{-1}s)\Phi)^{(n-1)}}_{\stensor^{n-1} W}\big| dk   
 < \infty.
\end{align*}
Thus  by Fubini's lemma, we have
\begin{align*}
&  \norm{d\Gamma(\chi_R)^{\frac{1}{2}}\Phi}^2 = \sum_{n=1}^\infty \biginner{\Phi^{(n)}}{\chi_R^{(n)}\Phi^{(n)}} \\
& = (2\pi)^{-\frac{3}{2}}\int_{\RR^3}d\bs\int   \hat\chi(\bs)  \sum_{n=1}^\infty
   \inner{(a(k)\Phi)^{(n-1)}}{(a(k-R^{-1}s )\Phi)^{(n-1)}}_{\stensor^{n-1} W} dk . 
\end{align*}
Thus \eqref{repr dgw} follows.
\end{proof}
%%%%%%%%%%%%%%%%%%%%%%%%%%%%%%%%%%%%%%%%%%%%%%%%%%%%%%%%%%%%%%%%
\begin{lem}\label{sp bd 1}
  Let $\{\Psi_m\}_{0<m<m_0} $ be normalized vectors in $\sH$ so that
\begin{enumerate}
\item[\rm (c1)] $\{ \Psi_m \}_{0<m<m_0} \subset \dom(\Nf^{\frac{1}{2}})$ and $\dstyle \sup_{0<m<m_0}\norm{\Nf^{\frac{1}{2}}\Psi_m}<\infty$,
\item[\rm (c2)] 
$$ \lim_{|\bs|\to 0} \sup_{0<m<m_0} \int \norm{a(k)\Psi_m-a(k-s)\Psi_m}^2 dk = 0,$$
where $s=(\bs,j)$ and $k-s = (\bk-\bs,j)$.
\end{enumerate}
Then  $\{\Psi_m\}_{0<m<m_0}$ satisfies that 
\begin{align}\label{zz614}
  \lim_{R\to\infty} \sup_{0<m<m_0} \norm{d\Gamma(\one-\chi_R)^{\frac{1}{2}}\Psi_m} = 0.
\end{align}
\end{lem}
%%%%%%%%%%%%%%%%%%%%%%%%%%%%%%%%%%%%%%%%%%%%%%%%%%%%%%%%%%%%%%%%
\begin{proof}
By Lemma \ref{lem of repr dGw} and $(2\pi)^{-\frac{3}{2}}\int_{\RR^3} \hat\chi(\bs) d\bs = \chi(0)=1$,
we have 
\begin{align*}
& \norm{d\Gamma( \one-\chi_R)^{\frac{1}{2}}\Psi_m}^2
 = \norm{\Nf^{\frac{1}{2}}\Psi_m}^2 - \norm{d\Gamma(\chi_R)^{\frac{1}{2}}\Psi_m}^2 \\
& = (2\pi)^{-\frac{3}{2}} \int_{\RR^3}d\bs  \int    \hat\chi(\bs)
   \inner{a(k)\Psi_m}{a(k)\Psi_m-a(k-R^{-1}s)\Psi_m} dk  \\
& \leq (2\pi)^{-\frac{3}{2}}\norm{\hat\chi}_{L^1}^{\frac{1}{2}} \norm{\Nf^{\frac{1}{2}}\Psi_m}
 \Big(\int_{\RR^3} d\bs |\hat\chi(\bs)|\int \norm{a(k)\Psi_m-a(k-R^{-1}s)\Psi_m}^2  dk \Big)^{\frac{1}{2}} \\
& \leq (2\pi)^{-\frac{3}{2}}\norm{\hat\chi}_{L^1}^{\frac{1}{2}} C \Big(\int_{\RR^3} |\hat\chi(\bs)|F_m(R^{-1}\bs) d\bs \Big)^{\frac{1}{2}},
\end{align*}
where $C=\sup_{0<m<m_0}\norm{\Nf^{\frac{1}{2}}\Psi_m}$ and 
\begin{align*}
 F_m(R^{-1}\bs) = \int \norm{a(k)\Psi_m-a(k-R^{-1}s)\Psi_m}^2 dk.  
\end{align*}
By condition (c1), we have  $F_m(R^{-1}\bs) \leq 4 C^2$ for all $m$. 
By condition (c2), for any $\vep>0$, there exists $M>0$ such that, for all $R>M$ and $|\bs|<R^{\frac{1}{2}}$,
 it holds that $\sup_{0<m<m_0} F_m(R^{-1}\bs)<\vep$.
Thus we have
\begin{align*}
& \sup_{0<m<m_0} \int_{\RR^3} |\hat\chi(\bs)|F_m(R^{-1}\bs) d\bs\\
& \leq \int_{|\bs|<R^{\frac{1}{2}}} |\hat\chi(\bs)| \vep d\bs 
   +  \int_{|\bs|>R^{\frac{1}{2}}} |\hat\chi(\bs)| 4C^2 d\bs 
    \leq \vep \norm{\hat\chi}_{L^1}   + 4C^2 \int_{|\bs|>R^{\frac{1}{2}}} |\hat\chi(\bs)| d\bs.
\end{align*}
Therefore
\begin{align*}
 \limsup_{R\to\infty} \Big( \sup_{0<m<m_0} \int_{\RR^3} |\hat\chi(\bs)|F_m(R^{-1}\bs) d\bs \Big)
 \leq  \vep \norm{\hat\chi}_{L^1}.
\end{align*}
Since $\vep>0$ is arbitrary, the lemma follows.
\end{proof}
%%%%%%%%%%%%%%%%%%%%%%%%%%%%%%%%%%%%%%%%%%%%%%%%%%%%%%%%%%%%%%%%
%\begin{df}
We extend Lemma \ref{sp bd 1}. %for later use. 
%%%%%%%%%%%%%%%%%%%%%%%%%%%%%%%%%%%%%%%%%%%%%%%%%%%%%%%%%%%%%%%%
%In order to prove Proposition \ref{photonloc}, we prepare the following lemma.
%%%%%%%%%%%%%%%%%%%%%%%%%%%%%%%%%%%%%%%%%%%%%%%%%%%%%%%%%%%%%%%%
\begin{lem}{\label{sp bd 2}}
  Let $\{\Psi_m\}_{0<m<m_0}$ be normalized vectors in $\sH$ so that
\begin{enumerate}
\item[\rm (a)] there exists $g\in W$ such that $\dstyle \sup_{0<m<m_0}\norm{a(k)\Psi_m} \leq |g(k)|$ for a.e.~$k$,
\item[\rm (b)] for any $0<\ep\ll 1$, 
    $$\lim_{|\bs|\to 0} \sup_{0<m<m_0} \int_{D_\ep^\mathrm{c}} \norm{a(k)\Psi_m-a(k-s)\Psi_m}^2 dk = 0,$$
where $k=(\bk,j), k-s=(\bk-\bs,j)$.
\end{enumerate}
Then \eqref{zz614} holds. 
\end{lem}
%%%%%%%%%%%%%%%%%%%%%%%%%%%%%%%%%%%%%%%%%%%%%%%%%%%%%%%%%%%%%%%%
\begin{proof}
From condition (a), the condition (c1) in Lemma \ref{sp bd 1} follows.
We shall  show (c2) in Lemma~\ref{sp bd 1}. By condition (a), we have
\begin{align}
& \sup_{0<m<m_0} \int \norm{a(k)\Psi_m-a(k-s)\Psi_m}^2 dk \notag \\
& \leq \sup_{0<m<m_0} \int_{D_\ep^\mathrm{c}} \norm{a(k)\Psi_m-a(k-s)\Psi_m}^2 dk
  + \int_{D_\ep} |g(k)|^2dk. \label{zz616}
\end{align}
By condition (b), the first term in \eqref{zz616} vanishes as $\bs\to 0$.
Thus 
\begin{align*}
0 \leq \limsup_{|\bs|\to 0} \sup_{0<m<m_0} \int \norm{a(k)\Psi_m-a(k-s)\Psi_m}^2 dk
  \leq \int_{D_\ep} |g(k)|^2 dk
\end{align*}
holds for all $\ep>0$. By the definition of $D_\ep$, the right-hand side of this 
inequality converges to zero as $\ep\to+0$.
Therefore, the condition (c2) in Lemma \ref{sp bd 1} is satisfied, and 
\eqref{zz614} holds.
\end{proof}
%%%%%%%%%%%%%%%%%%%%%%%%%%%%%%%%%%%%%%%%%%%%%%%%%%%%%%%%%%%%%%%%
%The spatial localization of photon for $\Phi_m$ is stated as follows.
%%%%%%%%%%%%%%%%%%%%%%%%%%%%%%%%%%%%%%%%%%%%%%%%%%%%%%%%%%%%%%%%
%%%%%%%%%%%%%%%%%%%%%%%%%%%%%%%%%%%%%%%%%%%%%%%%%%%%%%%%%%%%%%%%

We are in the position to prove Proposition \ref{photonloc}.
%%%%%%%%%%%%%%%%%%%%%%%%%%%%%%%%%%%%%%%%%%%%%%%%%%%%%%%%%%%%%%%% end proof
\begin{proof}[Proof of Proposition \ref{photonloc}:]
It is shown that 
$ \lim_{R\to\infty }\sup_{0<m<m_0} \norm{d\Gamma(\one-\chi_R)^{\frac{1}{2}}\Phi_m}^2 = 0$ implies 
\eqref{sploc} by \cite[IV.13]{ge00}. 
Hence 
it is sufficient to show that conditions (a) and (b) in Lemma \ref{sp bd 2}
are satisfied with $\Psi_m$ replaced by $\Phi_m$.
Proposition \ref{pdenb} yields that 
\begin{align*}
 \sup_{0<m<m_0} \norm{a(k)\Phi_m} 
 \leq C\frac{|\hat\varphi(\bk)|}{\ome(\bk)^{\frac{1}{2}}}(1+|\bk|),
 \qquad \text{a.e. } k 
\end{align*}
and the right-hand side above is square integrable in $k$ by (A2).
Thus condition (a) holds. Condition (b) is shown in Theorem \ref{econt}.
\end{proof}
%%%%%%%%%%%%%%%%%%%%%%%%%%%%%%%%%%%%%%%%%%%%%%%%%%%%%%%%%%%%%%%%
\section{Proof of the main theorem} \label{sec:pmt}
%In this section, we give the proof of Theorem \ref{MainThm}.
We show two general lemmas below. 
For a self-adjoint operator $A$, we denote the form domain of $A$ by $Q(A)$, 
and $(\, \cdot\,, A \,\cdot\, )$ denotes the quadratic
 form associated with $A$.
If $A$ is bounded from below, we set $E_0(A)=\inf\sigma(A)$.
For self-adjoint operators $A,B$, we denote
$A\geq B$ if and only if $Q(A)\subset Q(B)$ and $(\Psi,A\Psi)\geq (\Psi,B\Psi)$ for all $\Psi\in Q(A)$.
We use the following fact.
\begin{lem}\label{CrGS1}
  Let $A, A_j, j=1,2,\ldots$, be self-adjoint operators bounded from below such that
$  A_1 \geq A_2 \geq \ldots \geq A$. 
Assume that there exists a subspace $D \subset Q(A_1)$ such that 
$D$ is a form core for $A$ and 
$  \lim_{j\to\infty} (\Phi, A_j\Phi) = (\Phi, A\Phi)$ for $\Phi\in D$.
Then
$ \lim_{j\to \infty} E_0(A_j) = E_0(A)$.
\end{lem}
%%%%%%%%%%%%%%%%
\begin{proof}
 By the variational principle, we have  $E_0(A) \leq  E_0(A_j) \leq (\Phi,A_j\Phi)$ for any normalized $\Phi\in D$.
Since $E_0(A_j)$ is monotone decreasing in $j$, it has a limit as $j\to\infty$.
Since $D$ is a form core for  $A$, we have
\begin{align*}
  E_0(A) \leq \lim_{j\to\infty} E_0(A_j) \leq \inf_{\Phi\in D,\norm{\Phi}=1} (\Phi,A\Phi) = E_0(A).
\end{align*}
Therefore $E(A_j)\to E(A_0)$ as $j\to\infty$.
\end{proof}
%%%%%%%%%%%%%%%%

\begin{lem}\label{CrGS2}
  Let $A, A_j, j=1,2,\ldots$, be self-adjoint operators bounded from below such that
$  A_1 \geq A_2 \geq \ldots \geq A$. Assume that $\lim_{j\to\infty}E_0(A_j)=E_0(A)$.
Let $\Phi_j\in Q(A_j), j=1,2,\ldots$, be a normalized sequence such that 
\begin{align*}
  \inner{\Phi_j}{A_j\Phi_j} \leq E_0(A_j) + o(j^0),
\end{align*}
and $\Phi_j$ weakly converges to some $\Phi$ as $j\to\infty$.
Then $\Phi\in \dom(A)$ and 
\begin{align*}
  A\Phi = E_0(A)\Phi
\end{align*}
holds. In particular, if $\Phi\neq 0$, $\Phi$ is a ground state of $A$.
\end{lem}
%%%%%%%%%%%%%%%%
\begin{proof}
  Since $\Phi_j\in Q(A_j) \subset  Q(A)$, we have 
  \begin{align*}
 0 
 \leq (\Phi_j,(A-E_0(A))\Phi_j) 
  \leq (\Phi_j,(A_j-E_0(A))\Phi_j) 
 \leq E_0(A_j) - E_0(A) + o(j^0) 
 \to 0 
\end{align*}
as $j\to\infty$. Thus $\norm{(A-E_0(A))^{\frac{1}{2}}\Phi_j}\to 0$ as $j\to\infty$.
For any $\Psi\in Q(A)$, 
\begin{align*}
 \inner{(A-E_0(A))^{\frac{1}{2}}\Psi}{\Phi}
 = \lim_{j\to\infty} \inner{(A-E_0(A))^{\frac{1}{2}}\Psi}{\Phi_j} 
 = \lim_{j\to\infty} \inner{\Psi}{(A-E_0(A))^{\frac{1}{2}}\Phi_j} 
 = 0
\end{align*}
This implies that $\Phi\in Q(A)$ and $(A-E_0(A))^{\frac{1}{2}}\Phi=0$, and therefore
$\Phi\in\dom(A)$ and $(A-E_0(A))\Phi=0$.
\end{proof}
%%%%%%%%%%%%%%%%
%%%%%%%%%%%%%%%%%%%%%%%%%%%%%%%%%%%%%%%%%%%%%%%%%%%%%%%%%%%%%%%%
We need a bound to show the main theorem. 
\begin{lem}{\label{ineq:hh15}}
Assume (A1)--(A4) and $V\in V_\mathrm{conf}\cup V_\mathrm{rel}$. Then, for all $m\geq 0$,
\begin{align}
  \norm{|\bp|\Psi}^2+\norm{\Hfm\Psi}^2 \leq C(\norm{H_m\Psi}^2+\norm{\Psi}^2),
  \qquad \Psi\in \dom(H_m)   \label{rel0}
\end{align}
holds for some $C$ independent of $m\geq 0$.
\end{lem}
%%%%%%%%%%%%%%%%%%%%%%%%%%%%%%%%%%%%%%%%%%%%%%%%%%%%%%%%%%%%%%%%
\begin{proof}
In the case of $V\in V_\mathrm{conf}$, the lemma was proved by \cite{hh15}.
Since the proof for the case of $V\in V_\mathrm{rel}$ is similar, we briefly give an outline of the proof.
By the definition of $V_\mathrm{rel}$, there exist constants $0<a<1$ and $0<b$ such that
\begin{align}
  \norm{V\Psi}\leq a\norm{|\bp|\Psi}+b\norm{\Psi}, 
  \quad \Psi\in \dom(H_m). \label{rel3}
\end{align}
Set $H_0=|\bp-\bA(\bx)|+\Hfm$ and take an arbitrary $\Psi\in \Hfin$. It is shown that for an arbitrary $\epsilon>0$,
\begin{align}
 \norm{H_0\Psi}^2
 & \geq 
  (1-\epsilon) \norm{|\bp-\bA(\bx)|\Psi}^2 
  + (1-\epsilon) \norm{\Hfm\Psi}^2 - C_\epsilon \norm{\Psi}^2 \notag \\
 & \geq \frac{1-\epsilon}{1+\epsilon} (\norm{|\bp|\Psi}^2+\norm{\Hfm\Psi}^2)
  - C_\epsilon' \norm{\Psi}^2  \label{rel1}
\end{align}
with some constants $C_\epsilon$ and $C_\epsilon'$ (see \cite{hh15}).
Thus  by \eqref{rel3}, \eqref{rel1} and 
\begin{align}
 \norm{H_0\Psi}\leq \norm{H_m\Psi}+\norm{V\Psi}, \label{rel4}
\end{align}
 we have (\ref{rel0}) for all $\Psi\in\Hfin$.
Since $\Hfin$ is a core for $H_m$, the lemma follows by a limiting argument. 
\end{proof}

%%%%%%%%%%%%%%%%%%%%%%%%%%%%%%%%%%%%%%%%%%%%%%%%%%%%%%%%%%%%%%%%
Now we are in the position to prove the main theorem.
\begin{proof}[Proof of Theorem \ref{MainThm}:]
We can choose a subsequence $\{\Phi_{m_j}\}_j$ such that $m_j\downarrow 0$ as $j\to\infty$ 
and $\Phi_{m_j}$ weakly converges to some vector $\Phi_0\in \sH$.
Applying Lemmas \ref{CrGS1} and \ref{CrGS2} under the identifications:
$A = H$, $A_j = H_{m_j}$, $\Phi_j = \Phi_{m_j}$, 
$D = \Hfin$ and $\Phi=\Phi_0$, 
we can see that  $\Phi_0\in \dom(H)$ and 
\begin{align}\label{mainR}
  H\Phi_0 = E_0\Phi_0, \qquad E_0=\inf\sigma(H).
\end{align}
Now we shall show that $\Phi_{m_j}$ strongly converges to $\Phi_0$.
We first claim that the following bounds hold.  
\begin{align}
  & \sup_{j\in{\mathbb N}} \norm{|\bx| \Phi_{m_j}} < \infty,  \label{bd1} \\
  & \sup_{j\in{\mathbb N}} \norm{|\bp| \Phi_{m_j}} < \infty,  \label{bd2} \\
  & \sup_{j\in{\mathbb N}} \norm{\Hf \Phi_{m_j}} < \infty,    \label{bd3} \\ 
  & \sup_{j\in{\mathbb N}} \norm{\Nf^{\frac{1}{2}} \Phi_{m_j}} < \infty, \label{bd4}\\ 
  & \lim_{R\to\infty} \sup_{j\in{\mathbb N}} \norm{(\one-\Gamma_R) \Phi_{m_j}} = 0.  \label{bd5}
\end{align}
By assumption (A4), bound \eqref{bd1}  holds.
By  Lemma \ref{ineq:hh15} and $\norm{\Hf\Psi}\leq \norm{\Hfm\Psi}$, we have 
both bounds \eqref{bd2} and \eqref{bd3}.
Bound \eqref{bd4} is shown by  Corollary \ref{NN} and 
\eqref{bd5} by 
Proposition~\ref{photonloc}. 
From \eqref{bd1}--\eqref{bd5}, we have
\begin{align*}
 \sup_{j\in{\mathbb N}} \norm{(1-\chi_\ell)\Phi_{m_j}} = o(R^0), \qquad \ell=1,\ldots,5
\end{align*}
as $R\to\infty$,
where 
$\chi_1 = j(|\bx|/R)$,
$\chi_2 = j(|\bp|/R)$,
$\chi_3 = j(\Nf/R)$,
$\chi_4 = j(\Hf/R)$ and 
$\chi_5 = \Gamma_R$.
Here $j(\cdot)$ is the smooth function defined by \eqref{jfunction}. 
This fact implies that 
\begin{align}
& \sup_{j\in{\mathbb N}}\norm{ (1-\chi_1\chi_2\chi_3\chi_4\chi_5)\Phi_{m_j} } \notag \\
& \leq \sup_{j\in{\mathbb N}}\Big( \norm{ (1- \chi_1) \Phi_{m_j} }
  + \norm{ \chi_1 (1-\chi_2) \Phi_{m_j}}
  + \norm{ \chi_1 \chi_2 (1-\chi_3) \Phi_{m_j}} \notag \\
&\qquad  + \norm{ \chi_1 \chi_2 \chi_3(1-\chi_4) \Phi_{m_j} }
  + \norm{ \chi_1 \chi_2 \chi_3 \chi_4 (1-\chi_5 ) \Phi_{m_j} } \Big) \notag \\
& \leq \sup_{j\in{\mathbb N}} \sum_{\ell=1}^5 \norm{(1-\chi_\ell)\Phi_{m_j}} 
 \leq o(R^0).  \label{fin-bd}
\end{align}
Since $\chi_1 \chi_2 \chi_3 \chi_4 \chi_5$ is compact in $\sH$ for all $R>0$, 
$\chi_1 \chi_2 \chi_3 \chi_4 \chi_5\Phi_{m_j}$ strongly converges to 
$\chi_1 \chi_2 \chi_3 \chi_4 \chi_5\Phi_0$ as $j\to\infty$.
Thus  by  \eqref{fin-bd}, we have
\begin{align*}
 \norm{\Phi_0} 
& = \lim_{R\to\infty}\norm{\chi_1\chi_2\chi_3 \chi_4 \chi_5\Phi_0}
 = \lim_{R\to\infty}\lim_{j\to\infty}\norm{\chi_1\chi_2\chi_3 \chi_4 \chi_5\Phi_{m_j}} \\
& \geq \limsup_{R\to\infty}\limsup_{j\to\infty}(1-\norm{(1-\chi_1\chi_2\chi_3 \chi_4 \chi_5)\Phi_{m_j}}) 
 \geq \limsup_{R\to\infty} (1-o(R^0))  = 1.
\end{align*}
We conclude that $\Phi_{m_j}$ strongly converges to $\Phi_0$. In particular 
$\Phi_0\not=0$. 
By \eqref{mainR} 
$\Phi_0$ is a normalized ground state of $H$.
Then the proof is complete. 
\end{proof}
We give examples of the existence of the ground state. 
\begin{exa}
Suppose (A1) and (A2), and $V\in V_{\rm conf}$.
Then $H_m$ has the ground state for each $m>0$ by \cite{hh16}. 
In this case  (A3) and (A4) are satisfied. 
Then $H$ also has the ground state. 
\end{exa}

%%%%%%%%%%%%%%%%%%%%%%%%%%%%%%%%%%%%%%%%%%%%%%%%%%%%%%%%%%%%%%%% end proof

%%%%%%%%%%%%%%%% Acknowledgment
\vspace*{10pt}
\noindent\textbf{Acknowledgments:}
F. Hiroshima  thanks a kind hospitality of Aarhus university in Denmark and 
the International Network Program of the Danish Agency for Science,
Technology and Innovation. 
This work was supported by JSPS KAKENHI Grant Number JP16H03942
and 
JSPS KAKENHI Grant Number JP16K17612.

%%%%%%%%%%%%%%%% end

%%%%%%%%%%%%%%%%%%%%%%%%%%%%%%%%%%%%%%%%%%%%%%%%%%%%%%%%%%%%%%%% Reference
%\newpage 
%% table of contents\small \tableofcontents
\end{document}